\documentclass[journal]{IEEEtran}
\usepackage{amsmath,amssymb,bm}
\usepackage{booktabs}
\usepackage{graphicx}
\usepackage{float}
\usepackage{tikz}
\usetikzlibrary{positioning,arrows.meta,calc}
\usepackage{pgfplots}
\pgfplotsset{compat=1.17}
\usepgfplotslibrary{fillbetween}
\usepackage{url}
\usepackage{algorithm}
\usepackage{algpseudocode}
\usepackage{balance}

\newcommand{\jj}{\mathrm{j}}
\newcommand{\mM}{\bm{M}}
\newtheorem{proposition}{Proposition}

\begin{document}

\title{Differentiable Synthesis and Yield Optimization of\\
Cross-Coupled Resonator Filters}

\author{
\centering
Orion~Georgiou\\[0.5ex]
\itshape
Department of Electrical Engineering and Computer Science\\
University of California, Irvine
}

\maketitle

\begin{abstract}
Coupling-matrix synthesis of cross-coupled resonator filters is difficult when
the desired topology is noncanonical. Automatic differentiation removes the
need for problem-specific sensitivity expressions, but does not by itself overcome the nonconvex optimization landscape. This work develops a staged differentiable synthesis framework and identifies the importance of its initial objective through a controlled comparison that varies only the objective of the first optimization phase.
For tenth- and sixteenth-order
benchmarks, directly matching a frequency-sampled response under the
prescribed topology reaches the target from $0$ of $24$ and $0$ of $12$
random starts, respectively. Replacing the first-stage objective with
characteristic-polynomial matching increases the success rate to $18$ of
$24$ and $12$ of $12$ (exact paired $p=7.6\times10^{-6}$ and
$4.9\times10^{-4}$).
The observed difference is not explained by local conditioning and is
consistent with saturation of the bounded response residual far from the
solution. A differentiable front end converts symmetric
multiband specifications into target polynomials and, when applied end to end
to a fabricated dual-band filter specification, finds a realization with one
fewer cross-coupling than the published design. Finally, differentiating
through Monte Carlo tolerance simulations enables yield-driven design
centering, increasing the model-predicted yield from $50.9\%$ to
$67.0\pm0.5\%$ under an additive coupling-error model, an improvement of
$16.1$ percentage points.
\end{abstract}


\section{Introduction}

\IEEEPARstart{T}{he} coupling-matrix representation of coupled-resonator
filters~\cite{atia1972} provides a compact connection between a filter's
scattering response and the couplings required to realize it. For several canonical coupling
topologies, analytic synthesis procedures are available, and similarity
transformations can convert canonical solutions into certain practical
configurations~\cite{cameron1999,cameron2003,tamiazzo2005}.
For an arbitrary prescribed topology, however, no general analytic synthesis
is available. The resulting problem is therefore commonly formulated as a
nonlinear optimization over the coupling matrix~\cite{shang2012}.

Previous numerical approaches have used manually derived gradients,
eigenvalue-based objectives, and combinations of global and local
optimization~\cite{shang2012,amari2000,lamecki2004,nicholson2009}. These
methods can successfully synthesize practical filters, but they leave a
fundamental implementation difficulty: the derivatives must generally be
re-derived when the model or objective changes. This becomes increasingly
restrictive when the synthesis is extended beyond nominal response matching,
for example to include alternative objective functions or manufacturing
tolerances.

Automatic differentiation (AD)~\cite{baydin2018} removes this derivative
bottleneck, serving the same sensitivity role as the adjoint analysis long
used in microwave CAD~\cite{nikolova2004} while obtaining the derivatives from
the forward code itself. Modern
differentiable numerical frameworks can evaluate derivatives through the
matrix operations required by coupling-matrix models without manually deriving
and implementing problem-specific sensitivities. This makes it possible to
change objectives and extend the model while retaining gradients that remain
consistent with the forward computation.

However, the availability of exact gradients does not make arbitrary-topology
synthesis easy. In particular, directly optimizing a frequency-sampled
$S$-parameter objective under the prescribed topology remains highly nonconvex
and unreliable. The formulation of the optimization problem, particularly the
objective used in the initial phase, therefore becomes central to making
gradient-based coupling-matrix synthesis practical. We address this limitation with a staged differentiable synthesis framework.

The method directly optimizes the prescribed topology using a staged
objective: characteristic-polynomial matching is followed by
$S$-parameter refinement. A response-preserving rotation from an
unconstrained solution is also investigated as an alternative route, but the
direct route is simpler and is the recommended procedure. Controlled
experiments show that the choice of the initial objective is critical:
replacing direct frequency-sampled response matching with
characteristic-polynomial matching changes the synthesis problem from one in
which none of the tested starts reaches the target to one that succeeds
reliably on both tenth- and sixteenth-order benchmarks.

The framework also provides a differentiable front end that converts symmetric
multiband specifications into the characteristic polynomials required by the
synthesis procedure. Applied end to end to the \emph{specification} of a
fabricated dual-band filter, it finds a realization with one fewer
cross-coupling than the published design uses. Finally, because the forward model remains differentiable after the coupling
matrix is perturbed, gradients can also be propagated through Monte Carlo
tolerance simulations to perform yield-driven design centering.

The main contributions of this work are:
\begin{enumerate}
\item a controlled study of the objective-function choice in gradient-based
      coupling-matrix synthesis, showing that direct frequency-sampled
      response matching fails consistently under the prescribed topology
      while characteristic-polynomial matching succeeds reliably;

\item a staged differentiable synthesis framework that uses
characteristic-polynomial matching followed by topology-constrained
$S$-parameter refinement, together with a response-preserving rotation route
for comparison and initialization;

\item a differentiable specification front end that converts symmetric
      multiband filter requirements into the target characteristic
      polynomials required by the synthesis procedure;

\item end-to-end validation on published benchmarks and on the specification
      of a fabricated dual-band filter, whose \emph{specification} admits a
      realization with one fewer cross-coupling than the published design
      uses, while no such realization was found in $40$ constrained starts
      for the response realized by the published rounded coefficients
      (Section~\ref{sec:hardware}); and

\item an extension to yield-driven design centering through differentiation
      of Monte Carlo tolerance simulations.
\end{enumerate}

Section~\ref{sec:practice} provides practical guidelines, based on the
experimental findings, on objective and optimizer selection, the number of
seeds to use, and what to do when synthesis fails.

All experiments reported in this paper are performed on a single CPU core.
The accompanying software is released with the paper, with every
coupling matrix, target and topology mask, the seeds and
pseudorandom-number-generator (PRNG) keys that
generate the Monte Carlo draws, and the per-sample accept/reject vectors on
which the paired statistics are computed.\footnote{Repository URL to be
inserted at acceptance; the archive accompanying the submission is
self-contained.}

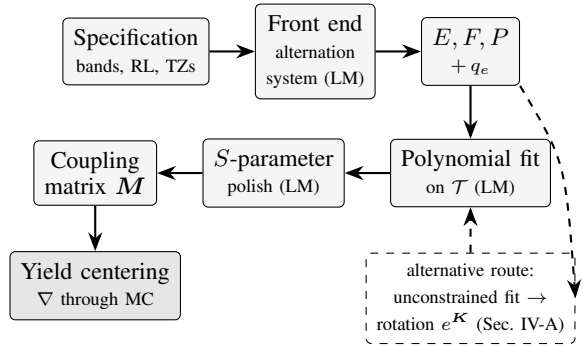
\begin{figure}[H]
\centering
\begin{tikzpicture}[
  font=\small,
  box/.style={draw, rounded corners=2pt, align=center, inner sep=4pt,
              minimum height=8mm, fill=gray!8},
  alt/.style={draw, rounded corners=2pt, align=center, inner sep=4pt,
              minimum height=8mm, fill=white, dashed},
  arr/.style={-{Stealth[length=2.5mm]}, thick},
  darr/.style={-{Stealth[length=2.5mm]}, thick, dashed}]
\node[box] (spec) {Specification\\ \scriptsize bands, RL, TZs};
\node[box, right=6mm of spec] (fe) {Front end\\ \scriptsize alternation\\ \scriptsize system (LM)};
\node[box, right=6mm of fe] (efp) {$E,F,P$\\ \scriptsize $+\,q_e$};
\node[box, below=7mm of efp] (sc) {Polynomial fit\\ \scriptsize on $\mathcal T$ (LM)};
\node[box, left=6mm of sc] (sd) {$S$-parameter\\ \scriptsize polish (LM)};
\node[box, left=6mm of sd] (out) {Coupling\\ matrix $\mM$};
\node[alt, below=6mm of sc] (rot) {\scriptsize alternative route:\\
   \scriptsize unconstrained fit $\to$\\ \scriptsize rotation $e^{\bm K}$
   (Sec.~\ref{sec:rotation})};
\node[box, below=6mm of out, fill=gray!20] (yield) {Yield centering\\ \scriptsize $\nabla$ through MC};
\draw[arr] (spec) -- (fe);
\draw[arr] (fe) -- (efp);
\draw[arr] (efp) -- (sc);
\draw[arr] (sc) -- (sd);
\draw[arr] (sd) -- (out);
\draw[arr] (out) -- (yield);
\draw[darr] (rot) -- (sc);
\draw[darr] (efp.south east) .. controls +(0.6,-1.0) .. (rot.east);
\end{tikzpicture}
\caption{Synthesis chain. Solid boxes form the recommended route; the dashed
box is the response-preserving rotation of Section~\ref{sec:rotation}, which
rejoins the direct route at the hard-masked polynomial-fit stage. Every
\emph{optimization} box obtains its derivatives by automatic differentiation:
the synthesis and front-end boxes are Levenberg--Marquardt (LM)
solves~\cite{marquardt1963} and use Jacobians, the yield centering box is Adam
on a Monte Carlo surrogate and uses gradients. The arrows themselves are
data flow. Spectral
factorization and Hurwitz-factor selection inside the front end are performed
outside the AD loop. No problem-specific derivative is derived by hand.}
\label{fig:pipeline}
\end{figure}

\section{Differentiable Coupling-Matrix Model}
\label{sec:model}
For an $N$-resonator network, let
$\mM\in\mathbb{R}^{N\times N}$ with $\mM=\mM^{T}$ denote the normalized
coupling matrix of a lossless, reciprocal network, and let
$q_{e1},q_{eN}>0$ denote the external quality factors at the input and output
ports. Following \cite{shang2012,hong2001}, the frequency-domain
coupling-matrix model is written as
\begin{equation}
[\bm A(s)] = [\bm q] + s[\bm I] - \jj[\mM],
\label{eq:A}
\end{equation}
where $[\bm q]$ is zero except $q_{11}=1/q_{e1}$, $q_{NN}=1/q_{eN}$, and
$s=\jj\Omega$, with $\Omega$ the normalized (dimensionless) low-pass
frequency.

For any given frequency, the reflection and transmission coefficients follow
from the inverse of $\bm A(s)$:
\begin{equation}
S_{11} = 1-\frac{2}{q_{e1}}[\bm A^{-1}]_{11},\qquad
S_{21} = \frac{2\,[\bm A^{-1}]_{N1}}{\sqrt{q_{e1}q_{eN}}}.
\label{eq:S}
\end{equation}
In the implementation, the required entries of $\bm A^{-1}$ are obtained by
solving the corresponding linear systems rather than explicitly forming the
inverse.

The same model can also be expressed in terms of the characteristic
polynomials $E$, $F$ and $P$. By Cramer's rule,
\begin{align}
E(s) &= \det \bm A(s), \nonumber\\
F(s) &= \det \bm A(s)
       -\frac{2}{q_{e1}}\operatorname{cof}_{11}\bm A(s), \\
\frac{P(s)}{\epsilon}
     &= \frac{2}{\sqrt{q_{e1}q_{eN}}}
        \operatorname{cof}_{1N}\bm A(s),
\label{eq:EFP}
\end{align}
so that
\begin{equation}
S_{11}(s)=\frac{F(s)}{E(s)},
\qquad
S_{21}(s)=\frac{P(s)}{\epsilon E(s)}.
\label{eq:SEFP}
\end{equation}
Both the sampled $S$-parameter response and the characteristic polynomials are
therefore functions of the same coupling matrix and external quality factors.

We implement these relations in JAX~\cite{jax2018}, using
\texttt{linalg.solve} to evaluate the scattering response and
\texttt{linalg.det} for the polynomial representation. Automatic
differentiation then supplies gradients or Jacobians of either representation
with respect to every coupling in $\mM$, the external quality factors, and any
differentiable upstream parameterization, so that changing the objective does
not require a new hand-derived sensitivity expression.

\subsection{Conventions Used Throughout}
\label{sec:conventions}
Table~\ref{tab:notation} collects the recurring symbols.

\begin{table}[htbp]
\caption{Recurring notation.}
\label{tab:notation}
\centering
\footnotesize
\begin{tabular}{ll}
\toprule
Symbol & Meaning\\
\midrule
$\mM$, $N$ & real symmetric coupling matrix; number of resonators\\
$q_{e1},q_{eN}$ & external quality factors at the two ports\\
$\Omega$, $s$ & normalized (dimensionless) frequency; $s=\jj\Omega$\\
$E,F,P,\varepsilon$ & characteristic polynomials and scale factor\\
$\widetilde F,\widetilde P$ & phase-adjusted real forms, Sec.~\ref{sec:frontend}\\
$\mathcal T$ & prescribed topology mask (allowed couplings)\\
$\bm r$, $\|\bm r\|_2^2$ & residual vector; the reported ``residual''\\
$\max|\Delta S|$ & response error, \eqref{eq:maxdS}\\
$\bm K(\bm\theta)$, $\bm Q$ & skew generator and orthogonal rotation, \eqref{eq:expK}\\
$B_+$, $n_b$ & positive-frequency passbands; reflection zeros in band $b$\\
$\sigma$, $\rho$, $\xi_{ij}$ & additive / multiplicative tolerance and variate\\
$T$, $\tau$, $\tau_\ell$ & acceptance threshold and surrogate smoothings\\
$\kappa_{\mathrm{eff}}$ & condition number on the complement of the nullspace\\
\bottomrule
\end{tabular}
\end{table}

Throughout this work, we use the following definitions:
\begin{itemize}
\item A \emph{residual} is always the squared Euclidean norm $\|\bm r\|_2^2$
      of the stated real residual vector, which is the quantity the LM
      iteration minimizes. Complex residuals enter $\bm r$ as their real and
      imaginary parts, stacked.
\item $\Delta S$ denotes the complex response difference from the reference,
      and
      \begin{multline}
      \max|\Delta S| = \max_{k}\,\max\bigl(
        |S_{11}(\jj\Omega_k)-S_{11}^{\mathrm{ref}}(\jj\Omega_k)|,\\
        |S_{21}(\jj\Omega_k)-S_{21}^{\mathrm{ref}}(\jj\Omega_k)|\bigr),
      \label{eq:maxdS}
      \end{multline}
      i.e.\ the maximum is taken over \emph{both} scattering parameters. The
      metric grid is $401$ uniform points on $\Omega\in[-2,2]$; independent
      validation grids are used in Section~\ref{sec:validation}.
\item A run is counted as a \emph{successful target-response reproduction}
      when $\max|\Delta S|<10^{-9}$ on that grid. We avoid calling this a
      global solution, which would suggest a proof of global optimality that
      we do not have. The threshold is applied to the response rather than to
      the entries because the coupling matrix is not unique: several matrices
      supported on the same topology can realize the same response.
\item \emph{Numerical environment:} all timings are for one core of an Intel
      Core~i5-1155G7 at $2.50$~GHz (processes pinned with \texttt{taskset};
      user time equals wall time in every run reported), JAX/jaxlib~$0.6.2$ on
      the CPU backend with \texttt{jax\_enable\_x64=True},
      \texttt{OMP\_NUM\_THREADS=1} and XLA configured with
      \texttt{--xla\_cpu\_multi\_thread\_eigen=false} and
      \texttt{intra\_op\_parallelism\_threads=1}. Reported wall times include
      JIT compilation. \texttt{jax.vmap} is used to \emph{vectorize} batches
      of independent evaluations; on this configuration it does not introduce
      parallel execution across cores. The comparison of
      Section~\ref{sec:bench16} is the one exception: it requires
      Octave for the reference implementation and was run on one core of an
      Intel Xeon at $2.80$~GHz, as stated there.
\end{itemize}

\section{What Makes the Optimization Work}
\label{sec:objective}

Automatic differentiation makes exact gradients available for any of the
objectives in Section~\ref{sec:model}, but their availability alone does not determine whether optimization reaches the target. This section isolates the effect of the objective using the tenth-order asymmetric dual-band benchmark (Example~A of~\cite{shang2012}, $N=10$, $25$ free
couplings) and the sixteenth-order quad-band benchmark (Example~B, $N=16$,
$24$ free couplings). The prescribed topology is enforced from the first
iteration in every case. 

The two benchmarks differ in ways beyond their order.
Example~A has an \emph{asymmetric} response, with unequal return-loss levels in
the two passbands and free self-couplings. Example~B is symmetric with
the diagonal constrained to zero. The results therefore do not depend on the symmetry assumptions of a particular prototype.

All experiments use the same initialization law,
$m_{ij}\sim\mathcal U(-0.7,0.7)$ on the allowed entries only. Comparisons also
use the same random initial coupling vectors, so differences in performance
are not attributable to different starting points.

\subsection{Step-size Sensitivity of the $S$-parameter Objective}
The natural objective is
\begin{multline}
\mathcal{L}_S(\mM)
=\sum_k
\Bigl[\,\bigl|S_{11}(\jj\Omega_k;\mM)-S_{11}^{\star}(\jj\Omega_k)\bigr|^2\\
+\bigl|S_{21}(\jj\Omega_k;\mM)-S_{21}^{\star}(\jj\Omega_k)\bigr|^2\,\Bigr],
\label{eq:LS}
\end{multline}
summed over the $401$-point metric grid. Table~\ref{tab:lrsweep} reports a
step-size sweep for Adam~\cite{kingma2015} over $64$ starts, $2000$
iterations, cosine decay to $2\%$ of the initial step.

The outcome depends strongly and non-monotonically on the step size. At
$10^{-3}$ no run comes within $10^{-3}$ of first-order stationarity: the
smallest gradient norm over the $64$ starts is $2.0$ and the median is
$16.7$, so the runs have not approached a stationary point and cannot be read
as poor local minima. As the step grows, the median gradient norm falls by more
than two orders of magnitude and the best residual by twenty-eight orders;
at $3\times10^{-1}$ the best and median residuals both deteriorate again,
though we observed no divergence or floating-point failure. At the best
setting, $10^{-1}$, an enlarged run of $512$ starts reaches
$5.3\times10^{-29}$ and $\max|\Delta S|=1.6\times10^{-15}$ in $9$ starts---a
success rate of $1.8\%$ at a cost of $673$~s.

Adam on $\mathcal L_S$ is therefore usable but unreliable and expensive, and
its behavior is dominated by a hyperparameter rather than by the geometry of
the problem. The next subsection tests whether a different objective improves convergence.

\begin{table}[htbp]
\caption{Adam on $\mathcal{L}_S$ under the prescribed topology, Example~A.
$64$ starts, $2000$ iterations, cosine decay. Residuals are
$\|\bm r\|_2^2$; ``conv.''\ counts starts with $\max|\Delta S|<10^{-9}$.}
\label{tab:lrsweep}
\centering
\footnotesize
\begin{tabular}{lcccc}
\toprule
Step & Best resid. & Med.\ resid. & Med.\ $\|\nabla\mathcal{L}_S\|_2$ & Conv.\\
\midrule
$10^{-3}$          & $3.1\times10^{-1}$  & $4.4\times10^{1}$ & $1.7\times10^{1}$ & 0\\
$3\times10^{-3}$   & $3.4\times10^{-2}$  & $2.1\times10^{1}$ & $3.1\times10^{0}$ & 0\\
$10^{-2}$          & $7.0\times10^{-4}$  & $2.0\times10^{1}$ & $4.7\times10^{-1}$ & 0\\
$3\times10^{-2}$   & $8.5\times10^{-13}$ & $1.2\times10^{1}$ & $1.7\times10^{-1}$ & 0\\
$\bm{10^{-1}}$     & $\bm{9.5\times10^{-29}}$ & $1.2\times10^{1}$ & $7.9\times10^{-2}$ & \textbf{2}\\
$3\times10^{-1}$   & $1.9\times10^{1}$   & $5.8\times10^{1}$ & $1.0\times10^{-1}$ & 0\\
\midrule
\multicolumn{5}{l}{\emph{$512$ starts at $10^{-1}$:} best $5.3\times10^{-29}$,
median $1.2\times10^{1}$,}\\
\multicolumn{5}{l}{$9/512$ converged, $\max|\Delta S|=1.6\times10^{-15}$,
$673$~s.}\\
\bottomrule
\end{tabular}
\end{table}

\subsection{A Polynomial-Value Objective}
\label{sec:polyobj}
The alternative is to match the characteristic polynomials \eqref{eq:EFP}
rather than the response. We evaluate $E$, $F$ and $P$ at $2\times24$ sample points, using $24$ equally spaced angles
$2\pi k/24$, $k=0,\dots,23$, on each of the two complex circles
$|s|=0.8$ and $|s|=1.4$. These circles bracket the passband
structure and provide samples on both sides of the relevant region. For each sampled quantity
$X\in\{E,F,P\}$, we use the residual
\begin{equation}
r = \frac{X-X^\star}{1+|X^\star|},
\label{eq:relres}
\end{equation}
which provides a mixed absolute--relative scaling. It behaves as a relative error when
$|X^\star|\gg1$ and as an absolute error when $X^\star$ passes through zero,
so that no sample point is either ignored or allowed to dominate.

The signs
and phases of the characteristic polynomials are not unique. We
therefore perform a short automatic search over the $\pm1$ and $\pm\jj$
conventions for $F$ and $P$ and select the convention that gives the smallest
polynomial residual relative to the target.

The eighth-order dual-band case in Section~\ref{sec:hardware} uses
$2\times20$ sample points with the same construction. The legacy Example-A
rotation script instead uses a single ring at $|s|=1.3$ with $24$ points and a
$201$-point response grid. All other results use the $2\times24$ configuration
described above.

\subsection{The Ablation}
\label{sec:ablation}
Table~\ref{tab:ablation} separates the contribution of the objective from that
of the optimizer. Every variant has the \emph{same} two-phase structure with
identical $300+300$ LM iteration budgets, starts from the same random coupling
vectors, and ends with the same LM phase on the $S$-parameter residuals. V1
and V3 therefore differ in exactly one thing: whether the first phase matches
the response or the characteristic polynomials. V2 and V4 prepend the same
$2000$-iteration Adam phase to V1 and V3 respectively, so the
stochastic-gradient contrast is budget-matched as well. Because the seeds are
shared, every comparison is tested by an exact paired McNemar test.

\begin{table}[htbp]
\caption{Ablation of the initial optimization objective. ``Jac.'' denotes Jacobian evaluations per seed in the LM phases;
Adam, where used, adds $2000$ gradient evaluations per seed. $p$ is the exact
paired McNemar test against the indicated row.}
\label{tab:ablation}
\centering
\scriptsize
\setlength{\tabcolsep}{3.5pt}
\begin{tabular}{llccccl}
\toprule
 & Phase 1 $\to$ phase 2 & Success & Jac. & Wall & s/succ. & $p$\\
\midrule
\multicolumn{7}{l}{\emph{Example A ($N=10$), 24 seeds}}\\
V1 & LM($S$)$\to$LM($S$)          & 0/24  & 461 & 91.6 s  & ---  & \\
V2 & Adam($S$)+V1                 & 0/24  & 443 & 213.5 s & ---  & $1.00$ vs V1\\
V3 & \textbf{LM(poly)$\to$LM($S$)}& \textbf{18/24} & 182 & 36.7 s & \textbf{2.0} & $\bm{7.6\times10^{-6}}$ vs V1\\
V4 & Adam(poly)+V3                & 16/24 & 179 & 65.2 s  & 4.1  & $0.69$ vs V3\\
\midrule
\multicolumn{7}{l}{\emph{Example B ($N=16$), 12 seeds}}\\
V1 & LM($S$)$\to$LM($S$)          & 0/12  & 562 & 176.4 s & ---  & \\
V2 & Adam($S$)+V1                 & 0/12  & 600 & 439.0 s & ---  & $1.00$ vs V1\\
V3 & \textbf{LM(poly)$\to$LM($S$)}& \textbf{12/12} & 314 & 72.6 s & \textbf{6.1} & $\bm{4.9\times10^{-4}}$ vs V1\\
V4 & Adam(poly)+V3                & 11/12 & 320 & 144.1 s & 13.1 & $1.00$ vs V3\\
\bottomrule
\end{tabular}
\end{table}

The initial objective decides the outcome: with the
optimizer, the seeds, the budget and the final phase all held fixed, replacing
the $S$-parameter first phase by the polynomial one takes the success rate
from $0/24$ to $18/24$ at $N=10$ ($p=7.6\times10^{-6}$) and from $0/12$ to
$12/12$ at $N=16$ ($p=4.9\times10^{-4}$). The polynomial route also uses
roughly half the Jacobian evaluations and a third to a half of the wall time.

The Adam phase did not improve success in these tests ($p=0.69$ and
$p=1.00$) while costing $2000$ extra gradient evaluations per seed and roughly
doubling the wall time. At these sample sizes a difference of this size would
not be detected, so we do not claim the two are equivalent; we omit the phase
from the recommended procedure of Section~\ref{sec:pipeline} on grounds of
cost, and we do not claim it degrades the result.

\subsection{Conditioning}
\label{sec:conditioning}
The difference in convergence is not explained by local conditioning. For
Example~B, the residual Jacobian has \emph{numerical rank $21$ and numerical
nullity $3$ at a relative tolerance of $10^{-12}$} at every point examined, for
both objectives. Its spectral condition number is therefore approximately
$10^{16}$ regardless of the objective and of the evaluation point. The raw condition number is consequently not informative for comparing
the two objectives.

Table~\ref{tab:kappa} reports the numerical rank at a relative tolerance of
$10^{-12}$, the corresponding nullity, and the effective condition number
\[
\kappa_{\mathrm{eff}}=\frac{\sigma_1}{\sigma_r},
\]
computed on the complement of the nullspace. These quantities are evaluated
both at the random initializations and at the \emph{successful} V3 endpoints.

\begin{table}[htbp]
\caption{Residual-Jacobian conditioning over the ablation seeds.}
\label{tab:kappa}
\centering
\footnotesize
\setlength{\tabcolsep}{4pt}
\begin{tabular}{lcccc}
\toprule
 & \multicolumn{2}{c}{Ex.~A (25 par., rank 25)} & \multicolumn{2}{c}{Ex.~B (24 par., rank 21)}\\
\cmidrule(lr){2-3}\cmidrule(lr){4-5}
Objective & at init & converged & at init & converged\\
\midrule
Polynomial    & $1.2\times10^{4}$ & $1.6\times10^{3}$ & $1.9\times10^{5}$ & $7.6\times10^{5}$\\
$S$-parameter & $1.2\times10^{3}$ & $9.2\times10^{1}$ & $3.0\times10^{2}$ & $3.9\times10^{1}$\\
\bottomrule
\end{tabular}
\end{table}

The conditioning is worse for the sixteenth-order benchmark. The effective
condition number for the polynomial objective increases from
$1.2\times10^{4}$ for Example~A to $1.9\times10^{5}$ for Example~B, consistent
with the attainable polynomial residual increasing from $10^{-28}$ to
$10^{-10}$. With only two filter orders, however, this does not establish a
trend with order, and the conditioning does not explain the difference between
the two objectives. The $S$-parameter Jacobian is better conditioned at both
orders and at both evaluation points, by one to four orders of magnitude, yet
the corresponding objective still fails to converge. Local conditioning
therefore does not explain the observed difference in optimization success.

\subsection{Redundancy in the Sixteenth-Order Topology}
\label{sec:redundant}
The three-dimensional nullspace is not numerical noise. Along each of the
three null directions, the response error grows \emph{quadratically} with
step size. The fitted slope of $\log\max|\Delta S|$ versus $\log\epsilon$ is
$2.00$ for all three directions, compared with $1.00$ for a generic
direction, confirming that they are tangent directions of the response map.
Moving a converged design a finite distance along each direction and
re-converging gives coupling matrices $0.20$--$0.25$ away in Euclidean norm
while reproducing the response to $\max|\Delta S|\le4.2\times10^{-15}$.
The prescribed Example-B topology therefore contains a three-parameter
continuous family of response-equivalent realizations and is redundant in the
sense of~\cite{lee2025}. This is a different situation from the enumeration of
\emph{isolated} solutions that homotopy-continuation frameworks
target~\cite{zeng2025homotopy}: here the solutions are not isolated points to
be counted but a continuum, along which the response is constant to machine
precision. This explains why the coupling values recovered for
this benchmark can differ substantially from the published matrix: the
optimization is selecting one point from a three-dimensional family rather
than recovering a unique set of coupling values.

This redundancy is not general, but is a property of the topology considered here. Where the residual Jacobian has locally constant rank, its nullity is
the dimension of the tangent space to the solution set and hence the local
dimension of the family of coupling matrices realizing the same response; the
quadratic growth and finite reconvergence reported above are the evidence that
this regularity holds at the Example-B solution. Over the seven topologies
used in this paper: the two literature benchmarks, the four stress-test
structures of Section~\ref{sec:stress}, and the folded dual-band topology of
Section~\ref{sec:hardware}, Example~B is the only one with nonzero nullity.
The other six have full-rank Jacobians both at the solution and at the random
initializations examined here, indicating locally unique coupling matrices.
The same check can be applied to a candidate topology using a single
singular-value decomposition.

The redundant directions also provide additional degrees of freedom that can
be used for a secondary objective. For example, the coupling values could be
selected to suit a particular physical realization, reduce sensitivity, or
improve tolerance robustness. In the latter case, the yield surrogate of
Section~\ref{sec:yield} could be optimized within the response-equivalent
family rather than re-centering the design and changing its nominal response.
We do not pursue this here because it requires an explicit parameterization of
the family, while the null directions provide only a local parameterization.

\subsection{Evidence Consistent with Saturation of the Response Residual}
\label{sec:saturation}
The remaining difference is global, not local. For a lossless reciprocal
network with real $\mM$ and $q_{e1},q_{eN}>0$, the scattering parameters
satisfy $|S_{11}|,|S_{21}|\le1$ on the real-frequency axis. Every entry of the
$S$ residual is therefore bounded by $2$, and
\begin{equation}
\|\bm r_S\|_2^2 \;\le\; 8N_{\mathrm{grid}} \;=\; 3208
\label{eq:bound}
\end{equation}
no matter how far $\mM$ is from the target. In contrast, the residuals constructed from $E$, $F$ and $P$ are
polynomial in the entries of $\mM$ and are unbounded.

Dilating the random initial coupling vectors by a factor of three changes the
median $\|\bm r_S\|_2^2$ by factors of $1.41$ at $N=10$ and $1.39$ at $N=16$. The corresponding median polynomial residual changes by factors of $3.3\times10^{5}$
and $3.1\times10^{7}$. Even before optimization,
$\|\bm r_S\|_2^2$ has a median value of $5.5\times10^{2}$ at $N=10$, already
within a factor of six of the bound in \eqref{eq:bound}.

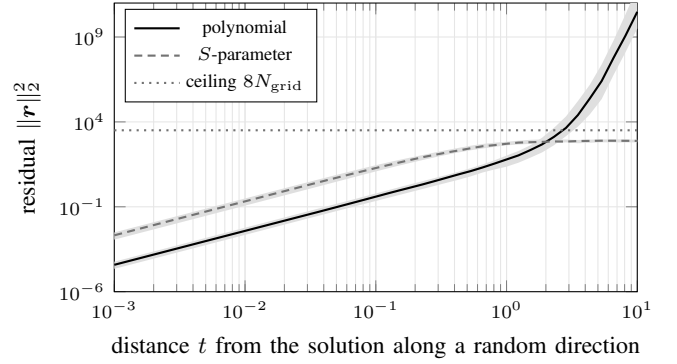
\begin{figure}[!t]
\centering
\begin{tikzpicture}
\begin{loglogaxis}[width=8.5cm, height=5.4cm,
  xlabel={distance $t$ from the solution along a random direction},
  ylabel={residual $\|\bm r\|_2^2$},
  xmin=1e-3, xmax=1e1, ymin=1e-6, ymax=1e11,
  legend style={font=\scriptsize, at={(0.02,0.98)}, anchor=north west},
  grid=both, grid style={gray!20}, tick label style={font=\scriptsize},
  label style={font=\small}]
\addplot[name path=pl, draw=none] table[x=t, y=pLo] {data/landscape.dat};
\addplot[name path=ph, draw=none] table[x=t, y=pHi] {data/landscape.dat};
\addplot[gray!25] fill between[of=pl and ph];
\addplot[name path=sl, draw=none] table[x=t, y=sLo] {data/landscape.dat};
\addplot[name path=sh, draw=none] table[x=t, y=sHi] {data/landscape.dat};
\addplot[gray!25] fill between[of=sl and sh];
\addplot[black, thick] table[x=t, y=pMed] {data/landscape.dat};
\addplot[black!55, thick, densely dashed] table[x=t, y=sMed] {data/landscape.dat};
\addplot[black!55, dotted, thick, domain=1e-3:1e1, samples=2] {3208};
\legend{,,,,,,polynomial, $S$-parameter, ceiling $8N_{\mathrm{grid}}$}
\end{loglogaxis}
\end{tikzpicture}
\caption{Both residuals along rays from a converged Example-A solution, median
over $24$ random unit directions (shaded band: 10th to 90th percentile). Near
the solution the two behave alike; far from it the $S$-parameter residual
flattens against its ceiling \eqref{eq:bound}, while the polynomial residual
keeps growing as a power law. This gap is what the ablation of
Table~\ref{tab:ablation} measures.}
\label{fig:landscape}
\end{figure}

Figure~\ref{fig:landscape} shows the same effect along rays
$\mM(t)=\mM^{\star}+t\bm d$ from a converged solution, in $24$ random unit
directions. For small $t$, both residuals increase together. Beyond
$t\approx0.5$, their behavior separates. Over the remaining factor of twenty
in distance, the $S$-parameter residual increases by a factor of $2.7$ and
approaches the bound in \eqref{eq:bound}, whereas the polynomial residual
increases by a factor of $2.4\times10^{9}$. Across the full sweep, the two
residuals span factors of $3.7\times10^{5}$ and $7.9\times10^{14}$,
respectively. A Gauss--Newton model formed from the nearly flat
$S$-parameter residual therefore provides little variation to drive the
iteration.

These results are evidence \emph{consistent with} saturation rather than a
proof of the mechanism. The dilation experiment is only a crude proxy for
distance from a solution set that is neither unique nor centered at the
origin, and the residual values alone do not determine the behavior of a
specific LM iteration. The bound in \eqref{eq:bound}, however, establishes
that the $S$-parameter residual has a limited dynamic range far from the
solution, whereas the polynomial residual does not. Together with the
ablation results, this is consistent with the larger dynamic range of the
polynomial objective contributing to its superior convergence.

\subsection{Sensitivity to the Sample-point Geometry}
\label{sec:rings}
Because the polynomial objective carries the method, its sample-point geometry
deserves scrutiny: the radii $|s|=0.8,1.4$ and the $24$ angles are choices, not
derivations. Table~\ref{tab:rings} varies both, using variant V3 and $12$ seeds
throughout.

\begin{table}[htbp]
\caption{Sensitivity of variant V3 to the complex sample-point geometry,
Example~A, $12$ seeds. $\kappa_2$ is the median condition number of the
polynomial Jacobian at the converged point.}
\label{tab:rings}
\centering
\footnotesize
\begin{tabular}{lccc}
\toprule
Sample points & Success & $\kappa_2(\bm J_{\mathrm{poly}})$ & Wall\\
\midrule
$|s|=0.8,1.4$; 24 angles (default) & 10/12 & $2.0\times10^{3}$ & 20 s\\
$|s|=0.6,1.2$; 24 angles & 7/12  & $2.5\times10^{3}$ & 11 s\\
$|s|=1.0,1.6$; 24 angles & 9/12  & $4.0\times10^{3}$ & 20 s\\
$|s|=0.9,1.1$; 24 angles & 11/12 & $1.3\times10^{3}$ & 13 s\\
$|s|=1.0$ only; 24 angles & 9/12 & $2.5\times10^{3}$ & 11 s\\
$|s|=0.8,1.4$; 12 angles & 8/12  & $3.1\times10^{3}$ & 11 s\\
$|s|=0.8,1.4$; 48 angles & 6/12  & $9.8\times10^{9}$ & 25 s\\
\bottomrule
\end{tabular}
\end{table}

Across this sweep the outcome varied little with the choice of radii, and a
single ring performed comparably to two rings. The number of sample points
mattered more: at the one denser setting tested, doubling the number of angles
to $48$ increased the Jacobian condition number by six orders of magnitude and
reduced the success rate. With $12$ seeds per setting and a single denser
configuration this is an observation from one sweep rather than a tested
rule, but it is consistent with the counting argument below.

The reason is a counting argument. Each of $E$, $F$, and $P$ is a polynomial
of degree at most $N$ and is therefore determined by $N+1$ coefficients.
Sampling at substantially more than $N$ points introduces residual rows that
are linearly dependent on the existing rows, up to numerical evaluation
error. These additional rows provide little new information but introduce
roundoff error and can worsen the numerical conditioning of the Jacobian.
Using a number of samples on the order of the polynomial degree is therefore
preferable. The $2\times24$ configuration used here provides sufficient
sampling for both $N=10$ and $N=16$ without the numerical penalty observed
with denser sampling.

The other free choice is the LM damping policy, which is conventional but
arbitrary. Re-running variant~V3 on Example~A with $\lambda_0$ varied from
$10^{-6}$ to $10^{-1}$, and with update factors from a gentle
$\times3/\times0.5$ to an aggressive $\times100/\times0.1$, gives $9$ or $10$
successes out of $12$ in every case, against $10$ for the policy of
Algorithm~1. The recommended route therefore has no delicate hyperparameter,
which is worth contrasting with Table~\ref{tab:lrsweep}, where the Adam step
size moves the outcome by twenty-eight orders of magnitude.

\subsection{Cost}
On the configuration of Section~\ref{sec:conventions}, a vectorized batch of
$64$ starts requires $81.3$~ms per Adam iteration on the $401$-point
$S$-parameter grid, compared with $42.0$~ms for the polynomial objective,
which uses a fixed $2\times24$ sample set independent of the response grid.
For a single start, one LM iteration, including the residual and
forward-mode Jacobian, requires $5.95$~ms for the $S$-parameter residual and
$6.07$~ms for the polynomial residual, a difference of less than $2\%$.
Thus, the two objectives have essentially the same per-iteration LM cost.
Combined with the Jacobian counts in Table~\ref{tab:ablation}, this makes the
polynomial objective less expensive overall.

\section{Differentiable Synthesis Procedure}
\label{sec:pipeline}

The ablation in Section~\ref{sec:ablation} determines the optimization
procedure. Given target polynomials $E^\star,F^\star,P^\star/\varepsilon$, a
topology mask $\mathcal T$, and $q_e$, the allowed couplings are first fitted
to the polynomial values using \eqref{eq:relres} with LM. The resulting
coupling matrix is then polished against the $S$-parameter response, also
using LM. No stochastic-gradient phase is used. Algorithm~\ref{alg:pipeline}
summarizes the procedure, including the LM damping policy, iteration budgets,
and initialization law.

\begin{algorithm}[!t]
\caption{Differentiable topology-constrained synthesis}
\label{alg:pipeline}
\begin{algorithmic}[1]
\Require targets $E^\star,F^\star,P^\star\!/\varepsilon$, $q_e$
(Sec.~\ref{sec:frontend}); topology mask $\mathcal T$
\Ensure coupling matrix $\mM$ supported on $\mathcal T$
\Statex \emph{LM policy (both phases):} damping $\lambda$ starts at
$10^{-3}$, $\times10$ on a rejected step, $\times0.3$ on an accepted step
(floor $10^{-14}$); at most $10$ damping trials per iteration; AD Jacobians,
forward mode wherever the residual vector is taller than the parameter vector.
\State \textbf{Init:} $m_{ij}\sim\mathcal U(-0.7,0.7)$ on $\mathcal T$;
independent seeds, $12$--$24$ of them
\State \textbf{P1:} LM on the relative $E,F,P$ residuals \eqref{eq:relres} at
$2{\times}24$ points ($|s|{=}0.8,1.4$), $\le300$ iterations; automatic search
over the $\pm1,\pm\jj$ phase conventions of $F$ and $P$
\State \textbf{P2:} LM on the complex $S$ residuals on the $401$-point grid
(forward-mode Jacobian), $\le300$ iterations $\to\mM$
\Statex \emph{Alternative entry (Sec.~\ref{sec:rotation}):} fit \emph{all}
off-diagonal couplings (plus diagonals iff the response is asymmetric) without
$\mathcal T$; then minimize the forbidden entries of
$e^{\bm K(\bm\theta)}\mM_0 e^{-\bm K(\bm\theta)}$ over $\binom{N-2}{2}$ angles;
hard-mask and continue at step~2, i.e.\ repeat P1 on the hard mask and
then run P2.
\end{algorithmic}
\end{algorithm}

\subsection{Response-Preserving Rotation onto the Topology}
\label{sec:rotation}

An alternative to constraining the topology from the first iteration is to
fit an unconstrained coupling matrix, a much easier problem that converges in
seconds to residuals of $10^{-24}$--$10^{-28}$ at $N=10$, and then transform
the resulting matrix to the prescribed topology without changing its response.

Let $\mM_0$ denote the unconstrained solution and let $\bm Q$ be an orthogonal
matrix that leaves the port coordinates fixed,
\begin{equation}
\bm Q\bm e_1=\bm e_1,\qquad \bm Q\bm e_N=\bm e_N .
\label{eq:portfix}
\end{equation}
Under $\mM(\bm\theta)=\bm Q\mM_0\bm Q^{T}$ the matrix $\bm A(s)$ of
\eqref{eq:A} transforms as $\bm A'(s)=\bm Q\bm A(s)\bm Q^{T}$, hence
$\bm A'(s)^{-1}=\bm Q\bm A(s)^{-1}\bm Q^{T}$, and by \eqref{eq:portfix} the
entries $[\bm A^{-1}]_{11}$ and $[\bm A^{-1}]_{N1}$ appearing in \eqref{eq:S}
are unchanged. The scattering response is exactly invariant.

We parameterize the transformation as
\begin{equation}
\bm Q(\bm\theta)=\exp\bigl(\bm K(\bm\theta)\bigr),
\label{eq:expK}
\end{equation}
where $\bm K$ is skew-symmetric and acts only on the $N-2$ internal resonator
coordinates. This gives $\binom{N-2}{2}$ independent parameters. The objective
is the squared magnitude of the forbidden entries of
$\bm Q\mM_0\bm Q^{T}$, together with the diagonal entries when the target
response is symmetric, with each undirected coupling counted once.
Automatic differentiation provides the Jacobian through both the matrix
exponential and the similarity transformation.

This is the reconfiguration problem studied in~\cite{wang2025}, where it is
solved using LM on the orthogonal group. Related formulations appear in
\cite{lee2025,wu2024}, in the matrix-completion approach of
\cite{zeng2024}, in the sequential-Householder framework of
\cite{wu2025householder}, and in the approximate Cayley
transform of~\cite{qian2025}. We claim no novelty for the problem itself. The
main structural difference from~\cite{wang2025} is the parameterization:
\eqref{eq:expK} uses a single global chart, whereas~\cite{wang2025} uses local
coordinates and applies a retraction at each iterate.

\begin{proposition}[$SO(N-2)$ suffices for zero-mask objectives]
\label{prop:so}
Let $\bm Q\in O(N-2)$ with $\det\bm Q=-1$, embedded as
$\mathrm{diag}(1,\bm Q,1)$, and let $\bm X=\bm Q\mM_0\bm Q^{T}$. Choose a
diagonal sign matrix $\bm D=\mathrm{diag}(d_i)$, $d_i\in\{\pm1\}$, acting on
the internal coordinates with $\det\bm D=-1$. Then
$\bm D\bm Q\in SO(N-2)$, it satisfies \eqref{eq:portfix}, and
$(\bm D\bm Q)\mM_0(\bm D\bm Q)^{T}=\bm D\bm X\bm D$ has entries
$d_id_j X_{ij}$. Hence $|(\bm D\bm X\bm D)_{ij}|=|X_{ij}|$ for all $i,j$: the
zero pattern and the squared forbidden-entry objective are identical.
Restricting the search to $SO(N-2)$ therefore loses no zero-mask solution.
\end{proposition}

Only the \emph{signs} of the retained couplings differ between the two
components, so a formulation that constrains coupling signs, rather than
their support, would need the reflection component as well.

\subsection{Comparison With the Published Reconfiguration Method}
\label{sec:bench16}

Because \cite{wang2025} addresses the same reconfiguration problem and provides
a runnable implementation, we compare the two methods directly. The results
are summarized in Table~\ref{tab:bench16}.

\emph{Protocol.} Both methods minimize the identical objective
$\|\bm W\odot(\bm Q^{T}\mM_0\bm Q)\|_F^2$ with
$\bm Q=\mathrm{diag}(1,\bm U,1)$, use the same $200$ random initial
orthogonal matrices $\bm U_0=\exp(\bm K)$, and are limited to $300$
iterations. The reference implementation is used as published, with one
documented modification: an option was added to supply $\bm U_0$ so that the
two methods can be evaluated from identical initial conditions. The released
implementation otherwise initializes from the identity or from its own random
draw. The upstream commit and the patch are included in the accompanying
archive.

Table~\ref{tab:bench16} declares a run successful when its final objective
falls below $10^{-12}$. Recomputing at $10^{-20}$ moves a single entry, our
method on cm8, from $160$ successes to $159$; every reference count and every
iteration median is unchanged, and the cm8 $p$-value moves from
$2.1\times10^{-3}$ to $1.4\times10^{-3}$. The comparison therefore does not
depend on where the threshold is placed. Both counts are recorded for every
case in the archive. Because the initial conditions are paired, we compare
the success outcomes using an exact paired McNemar test. The first three rows
are the reference repository's test cases; the last two are the rotation
subproblems from Examples~A and~B in this paper, expressed in the same format.

Wall times are reported in the accompanying archive but not compared here
because the reference implementation runs in Octave whereas ours runs in
JAX~0.10.2. The latter differs from the JAX~0.6.2 environment used elsewhere
in this paper because the comparison requires a machine with Octave. The
iteration and success counts therefore provide the direct comparison.

\begin{table}[htbp]
\caption{Response-preserving reconfiguration comparing the reference
implementation of~\cite{wang2025} with the matrix-exponential parameterization
of Section~\ref{sec:rotation}. Both methods use the same objective,
$200$ initial orthogonal matrices, and a $300$-iteration cap. $p$ is the exact
paired McNemar test.}
\label{tab:bench16}
\centering
\footnotesize
\setlength{\tabcolsep}{4pt}
\begin{tabular}{lccccc}
\toprule
 & \multicolumn{2}{c}{Success / 200} & \multicolumn{2}{c}{Median iterations} & \\
\cmidrule(lr){2-3}\cmidrule(lr){4-5}
Case ($N$, inner) & \cite{wang2025} & Ours & \cite{wang2025} & Ours & $p$\\
\midrule
cm4 \ \ (6, 4)    & 200 & 200 & \textbf{9}  & 13  & $1.00$\\
cm8 \ \ (10, 8)   & \textbf{182} & 160 & \textbf{28} & 110 & $2.1\times10^{-3}$\\
cm10 (12, 10)     & \textbf{121} & 91  & \textbf{49} & 149 & $4.2\times10^{-3}$\\
Ex.~A (10, 8)     & 99  & 103 & \textbf{69} & 91  & $0.74$\\
Ex.~B (16, 14)    & 0   & 0   & 300 & 300 & $1.00$\\
\bottomrule
\end{tabular}
\end{table}

The comparison shows a clear limitation of the single global chart. On the
reference method's own benchmarks, the retraction-based formulation of
\cite{wang2025} performs significantly better, and the difference increases
with the number of internal resonators. The methods are tied at four internal
coordinates, but the reference method achieves $182$ versus $160$ successes at
eight ($p=2.1\times10^{-3}$) and $121$ versus $91$ at ten
($p=4.2\times10^{-3}$), with median iteration counts smaller by factors of
$3.9$ and $3.0$, respectively. On the Example-A rotation problem, no
significant difference is detected ($99$ versus $103$, $p=0.74$). An earlier
20-start experiment appeared to favor the global chart, but that difference
does not persist with the larger sample. Most importantly, neither method
reconfigures the sixteenth-order quad-band problem in any of the $200$ starts;
both stall at a forbidden-coupling residual of $4$--$6\times10^{-2}$.

The difference on the reference benchmarks is also reflected in the geometry
of the parameterization. A single global chart must represent the
transformation with one skew-symmetric generator $\bm K$, and the required
generator becomes larger as the reconfiguration becomes more difficult. Among
the converged runs, the median $\|\bm K\|_F$ is $8.1$ for cm8 and $15.9$ for
cm10, compared with $17.9$ and $25.6$ for the failed runs. At the failed
points, the median $\kappa_2$ of the rotation Jacobian is
$1.5\times10^{3}$ and $1.1\times10^{4}$, compared with
$3.1\times10^{2}$ for the successful runs. These results are consistent with
the single chart becoming poorly conditioned as the required transformation
moves farther from the identity. The retraction-based method of
\cite{wang2025} instead recenters the local coordinates at each iterate,
avoiding accumulation in a single $\bm K$. For practical response-preserving
reconfiguration, the tested method of~\cite{wang2025} is therefore preferable
to the chart of \eqref{eq:expK} on these benchmarks; the other cited
parameterizations~\cite{lee2025,wu2024,zeng2024,wu2025householder,qian2025}
were not compared here and warrant a separate evaluation.

\subsection{Why the Rotation Route Fails at Sixteenth Order}
\label{sec:whyfail}
Both implementations fail on the Example-B rotation problem, whereas direct synthesis succeeds from every seed. The available evidence points to a limitation of the matrix passed to the rotation stage rather than to the construction or numerical stability of the rotation itself.

Constructing a port-fixing rotation is not the limiting step. Such a
transformation preserves every port moment
$\bm e_i^{T}\mM^{k}\bm e_j$ with
$i,j\in\{1,N\}$. When the block-Krylov space generated by
$\{\bm e_1,\bm e_N\}$ spans $\mathbb R^N$, as it does for these cases, those
moments determine $\mM$ up to a port-fixing orthogonal transformation. This is
the block-Lanczos argument: Gram--Schmidt on
$\bm e_1,\bm e_N,\mM\bm e_1,\mM\bm e_N,\dots$ needs only the inner products
$\langle\mM^{a}\bm e_i,\mM^{b}\bm e_j\rangle=\bm e_i^{T}\mM^{a+b}\bm e_j$,
so the basis $\bm V$ and the compression $\bm V^{T}\mM\bm V$ are fixed by the
moments alone; equal moments therefore give
$\mM'=(\bm V'\bm V^{T})\mM(\bm V'\bm V^{T})^{T}$, and $\bm V'\bm V^{T}$
fixes $\bm e_1,\bm e_N$ because $\bm V$ and $\bm V'$ share their first two
columns. The connecting transformation can then be constructed explicitly as
\begin{equation}
\bm Q=\bm V_{\mathrm{tgt}}\bm V_{\mathrm{src}}^{T},
\end{equation}
where the two matrices contain the corresponding block-Krylov orthonormal
bases. For matrix pairs that lie on the same orbit, this construction recovers
the connecting $\bm Q$ directly and preserves the port vectors to machine
precision.

The construction does amplify, but only by an order-one factor. Perturbing an
exactly supported matrix by a random symmetric $\bm E$ scaled so that
$\|\bm E\|_{\max}=\epsilon$, and reconstructing, displaces the result by
$\|\Delta\mM\|_{\max}/\|\bm E\|_{\max}$ with a median over five trials of
$0.96$--$1.30$ at $N=10$ and $1.79$--$2.23$ at $N=16$, for $\epsilon$ between
$10^{-10}$ and $10^{-4}$; the largest single trial gives $4.76$. Measured
instead in the Frobenius norm on both sides the median ratios are $1.05$--$1.33$
throughout. Five trials at each setting are too few to characterize the
conditioning of the map, but no amplification large enough to explain the
$O(10^{-1})$ orbit discrepancy was found: the worst observed factor applied to
the largest $\epsilon$ tested gives $4.8\times10^{-4}$, still roughly $300$
times smaller than $0.136$, and the discrepancy would need a factor of nearly
$1400$ at that $\epsilon$. The observed
failure of the rotation route therefore does not appear to result from
numerical instability in this reconstruction.

The same construction shows that the Stage-A matrix $\mM_0$ and the published
Example-B matrix $\mM_B$ do not lie on the same orbit. Their port moments
differ, and the transformation constructed from their block-Krylov bases gives
\begin{align*}
\bigl\|\bm Q\mM_0\bm Q^{T}-\mM_B\bigr\|_{\max}
&:=\max_{i,j}\bigl|\bigl(\bm Q\mM_0\bm Q^{T}-\mM_B\bigr)_{ij}\bigr|\\
&\;=0.136,
\end{align*}
which is comparable to the coupling values themselves. Since the port moments
determine a full block-Krylov matrix up to a port-fixing rotation, no
port-fixing orthogonal transformation maps $\mM_0$ to this particular target.
Stage~A nevertheless matches the target polynomials to
$1.6\times10^{-10}$. As shown in Section~\ref{sec:conditioning} and
Table~\ref{tab:validate}, however, the polynomial residual at this order
constrains the coupling matrix only weakly. Thus, a small polynomial residual
can correspond to a matrix that is $O(10^{-1})$ away in its entries from the
published realization.

A rotation search starting from $\mM_0$ is confined to the orbit of $\mM_0$.
We cannot determine from the moment comparison whether that orbit contains
\emph{some other} supported matrix with the desired response. The search
results nevertheless provide no evidence that it does: our exponential chart
fails in all $80$ starts in Table~\ref{tab:bench}, and the two implementations
in Section~\ref{sec:bench16} both fail in all $200$ matched starts. Direct
synthesis is not subject to this orbit constraint because it re-optimizes the
couplings themselves. The hard-masked polish used by Route~R is likewise able
to modify the coupling matrix after the rotation stage, which explains why
Route~R can still finish at this order even though its rotation stage does not
reach the prescribed topology.

This issue is not specific to Example~B. Whenever the block-Krylov space is
full and a candidate target matrix is available, the same construction can be
used as a pre-check: form the two block-Krylov bases, construct
$\bm Q=\bm V_{\mathrm{tgt}}\bm V_{\mathrm{src}}^{T}$, and evaluate
$\|\bm Q\mM_0\bm Q^{T}-\mM_{\mathrm{tgt}}\|$. A value substantially above the
reconstruction error indicates that the two matrices lie on different
port-fixing orbits and that no rotation search starting from $\mM_0$ can reach
that particular target.

The rotation problem also becomes increasingly constrained with filter order.
There is one residual for each forbidden entry and one degree of freedom for
each rotation angle. At $N=10$, this gives $30$ constraints and $28$ angles;
at $N=16$, there are $112$ constraints and $91$ angles. Feasible instances can
still exist because the target may lie on the orbit, but the larger problem
requires satisfying substantially more constraints in a higher-dimensional
search space.

\subsection{Choosing Between the Direct and Rotation Routes}
\label{sec:route}

We recommend the direct route on the basis of its simplicity, not because it
shows a demonstrated performance advantage over the rotation route. Across
the matched starts in Table~\ref{tab:bench}, the direct route succeeds
somewhat more often at both orders, but the differences are not statistically
resolved at these sample sizes. It also requires none of the additional
rotation machinery: there is no unconstrained fit, no parameterization on
$\mathrm{SO}(N-2)$, no search over $28$ or $91$ rotation angles, and no
separate rotation stage. The direct route therefore provides the simpler
pipeline without sacrificing the observed success rates.

This result also changes the role of the unconstrained-fit-plus-rotation
decomposition at higher order. We initially expected the decomposition to
become necessary as the filter order increased. On these benchmarks, it is not
necessary, although the rotation stage remains useful in some settings,
particularly as an initializer for the subsequent hard-masked optimization.

The rotation should not, however, be treated as a step that must reach the
prescribed topology. At $N=16$, neither implementation reaches the topology in
any of the tested starts. A pipeline that depends on successful rotation will
therefore fail at this order. When used only to provide an initial matrix for
the hard-masked polish, however, the rotation route remains viable, as shown
for Route~R in Table~\ref{tab:bench}.

The rotation formulation is also useful as a diagnostic. Its residual measures
how closely the response-preserving orbit of $\mM_0$ approaches the prescribed
support. A large residual can therefore indicate that the topology is poorly
matched to the response, as discussed in Section~\ref{sec:stress}. At high
order, however, Section~\ref{sec:whyfail} shows that the same observation can
also arise because $\mM_0$ lies on a different response-preserving orbit from
the available target realization. The rotation residual should therefore be
interpreted as a diagnostic of the chosen starting matrix and topology, not as
a definitive test of whether the topology can represent the response.

\begin{figure}[!t]
\centering
\begin{tikzpicture}[scale=0.78, every node/.style={transform shape},
  res/.style={circle, draw, thick, minimum size=6.5mm, inner sep=0pt,
              font=\small},
  ml/.style={thick},
  cx/.style={thick, dashed}]
\node[res] (n1) at (0,0) {1};
\node[res] (n2) at (0.4,1.5) {2};
\node[res] (n3) at (1.7,1.9) {3};
\node[res] (n4) at (1.7,0.4) {4};
\node[res] (n5) at (1.9,-1.1) {5};
\node[res] (n6) at (3.4,-1.1) {6};
\node[res] (n7) at (3.6,0.4) {7};
\node[res] (n8) at (3.6,1.9) {8};
\node[res] (n9) at (5.0,1.5) {9};
\node[res] (n10) at (5.2,0) {10};
\draw[ml] (n1)--(n2) (n2)--(n3) (n3)--(n4) (n4)--(n5) (n5)--(n6)
          (n6)--(n7) (n7)--(n8) (n8)--(n9) (n9)--(n10);
\draw[cx] (n1)--(n3) (n1)--(n4) (n3)--(n8) (n4)--(n7) (n7)--(n10)
          (n8)--(n10);
\draw[thick,-{Stealth}] (-1.0,0)--(n1); \node at (-1.25,0.35) {\scriptsize P1};
\draw[thick,-{Stealth}] (6.2,0)--(n10); \node at (6.45,0.35) {\scriptsize P2};
\end{tikzpicture}
\caption{Topology of the tenth-order asymmetric dual-band benchmark
(Example~A of~\cite{shang2012}). Solid: mainline couplings; dashed:
cross-couplings. Self-couplings (nonzero diagonal) are present but not drawn.}
\label{fig:topoA}
\end{figure}
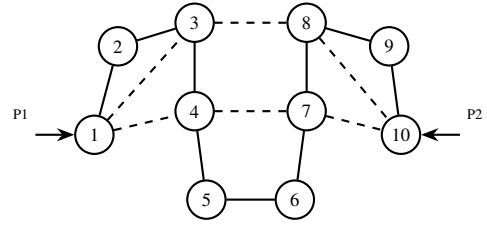

\section{Specification Front End}
\label{sec:frontend}

The synthesis procedure in Section~\ref{sec:pipeline} takes the target
characteristic polynomials as input. A designer instead specifies passband
edges $[a_b,b_b]$, return-loss levels $L_{Rb}$, and transmission-zero (TZ)
locations. The front end converts these specifications into the target
polynomials $E$, $F$, $P$ and the external quality factor $q_e$ required by the
synthesis procedure.

\subsection{Real Phase-Adjusted Polynomials}
\label{sec:phasepoly}

Throughout this section, the response is even in $\Omega$, and the transmission
zeros lie on the imaginary axis in conjugate pairs. It is therefore convenient
to work with the phase-adjusted real polynomials
\begin{equation}
\widetilde F(\Omega)\;\propto\;F(\jj\Omega),
\qquad
\widetilde P(\Omega)\;\propto\;P(\jj\Omega),
\label{eq:phaseadj}
\end{equation}
where the proportionality constants are unimodular factors chosen so that
$\widetilde F$ and $\widetilde P$ are real. For reflection zeros at
$\pm z_r$ and transmission zeros at $\pm\jj t_k$, this gives
\begin{equation}
\widetilde F(\Omega)
=
c_F\prod_{r=1}^{N/2}\bigl(z_r^2-\Omega^2\bigr),
\qquad
\widetilde P(\Omega)
=
c_P\prod_{k=1}^{n_z}\bigl(t_k^2-\Omega^2\bigr),
\label{eq:Ftilde}
\end{equation}
where the product for $\widetilde F$ runs over the $N/2$ positive-frequency
reflection zeros, giving a polynomial of degree $N$ in $\Omega$. Without using
the even symmetry, the same polynomial could instead be written as a product
over all $N$ signed reflection zeros. The reduced form in
\eqref{eq:Ftilde} is the one used in the implementation.

The unimodular factors in \eqref{eq:phaseadj} are the same
$\pm1,\pm\jj$ phase conventions considered by the automatic phase search in
Section~\ref{sec:polyobj}.

\subsection{Alternation System}
Writing the characteristic function as
\begin{equation}
k(\Omega)=\varepsilon\,
\frac{\widetilde F(\Omega)}{\widetilde P(\Omega)},
\label{eq:kfun}
\end{equation}
the prescribed TZs determine $\widetilde P$ and the reflection zeros determine
$\widetilde F$. For an equiripple response, $k^2$ must attain the same
prescribed level at both edges of each passband and at each interior ripple
extremum. In positive-frequency band $b$ that level is
\begin{equation}
k_b^2=\bigl(10^{L_{Rb}/10}-1\bigr)^{-1}.
\label{eq:ripple}
\end{equation}
Let $B_+$ denote the number of \emph{positive-frequency} passbands (half the
total number of passbands for a symmetric response) and let $n_b$ reflection
zeros be assigned to band $b$, so that
\begin{equation}
\sum_{b=1}^{B_+} n_b=\frac{N}{2}.
\label{eq:zerocount}
\end{equation}
Band $b$ then has $n_b-1$ interior extrema $\mu_j$, and the alternation
conditions are
\begin{equation}
k^2(\Omega)=k_b^2
\label{eq:level}
\end{equation}
at both band edges and at every $\mu_j$, together with the stationarity
condition
\begin{equation}
\widetilde F'(\mu_j)\widetilde P(\mu_j)
-\widetilde F(\mu_j)\widetilde P'(\mu_j)=0 .
\label{eq:stationary}
\end{equation}
The construction is related to the Remez-like multiband prototype synthesis
of~\cite{macchiarella2013} and the recent high-order formulation of
\cite{zhao2026}. Here, however, the reflection-zero locations $z_r$, the
interior extrema $\mu_j$, and the scale factor $\varepsilon$ are solved
simultaneously as a single nonlinear least-squares problem. LM provides the
optimization step, while automatic differentiation supplies the Jacobian.
There is no exchange iteration.

\emph{Counting.} With all band edges, return-loss levels and TZ locations
fixed, the unknowns are the $N/2$ reflection zeros, the
$\sum_b(n_b-1)=N/2-B_+$ interior extrema and $\varepsilon$, i.e.\
$N-B_++1$ unknowns; the conditions are $2B_+$ edge levels,
$N/2-B_+$ extremum levels and $N/2-B_+$ stationarity conditions, i.e.\ $N$
conditions. The system is therefore overdetermined by exactly
\begin{equation}
N-(N-B_++1)=B_+-1
\label{eq:overconstraint}
\end{equation}
conditions. Following~\cite{shang2012}, we remove this overconstraint by
allowing $B_+-1$ transmission-zero locations or ripple levels to vary. For the
quad-band example below, $N=16$ and $B_+=2$, so one condition is freed and the
resulting system is square, with $16$ unknowns and $16$ equations.

\subsection{Implementation Details}
\emph{1) Order-preserving parameterization.} The squared alternation equations
otherwise admit nonphysical solutions in which reflection zeros leave their
assigned passband or collapse onto a band edge. We prevent
this by parameterizing the zeros through unconstrained increments
$u_1,\dots,u_{2n_b}$. After a softplus transformation and cumulative
normalization, they satisfy
\begin{equation}
a_b<z_1<\mu_1<z_2<\cdots<z_{n_b}<b_b ,
\label{eq:order}
\end{equation}
while $\varepsilon=e^{u_0}$ enforces $\varepsilon>0$. The optimization is
therefore restricted to the physically admissible region throughout.

\emph{2) Scale-free residuals.} When a transmission zero lies close to a passband, $\widetilde P(\Omega)$ can
vary by up to ten orders of magnitude across the frequency range, making the
direct residuals poorly scaled. In these cases, we use the logarithmic form of
the level condition,
\begin{equation}
\log\bigl(\varepsilon^2\widetilde F^{\,2}\bigr)
-\log\bigl(k_b^2\widetilde P^{\,2}\bigr),
\label{eq:logres}
\end{equation}
and write the stationarity condition as the logarithmic derivative
\begin{equation}
\frac{\widetilde F'}{\widetilde F}
-\frac{\widetilde P'}{\widetilde P}.
\label{eq:logder}
\end{equation}
These residuals depend on the relative variation of the polynomials rather
than their absolute magnitudes. By contrast, bounded normalized residuals such
as $(a-b)/(|a|+|b|)$ saturate at $\pm1$ with vanishing gradients and can create
spurious LM attractors; this behavior is observed in the quad-band example.
The dual-band case of Section~\ref{sec:hardware} is sufficiently well scaled
that the direct forms in \eqref{eq:level}--\eqref{eq:stationary} are adequate,
and these are used in the released dual-band script.

\emph{3) Recovery of $E(s)$.} Once $\widetilde F$, $\widetilde P$, and $\varepsilon$ have been determined,
$E(s)$ is recovered outside the AD loop by classical spectral factorization,
\begin{equation}
|E(\jj\Omega)|^2
=
\widetilde F^{\,2}(\Omega)
+\frac{\widetilde P^{\,2}(\Omega)}{\varepsilon^2}.
\label{eq:specfact}
\end{equation}
The roots are computed from a companion-matrix eigenvalue problem, and the
Hurwitz factor is selected. The external quality factor $q_e$ then follows
from the coefficient of $s^{N-1}$ in $E$, as in~\cite{shang2012}.

The formulation assumes real $\widetilde F$ and $\widetilde P$ and an ordered
real-root parameterization, so it does not cover complex transmission zeros
used for group-delay equalization. Extending the formulation would require
modulus-squared residuals together with a conjugate-pair parameterization of
the zero locations; this extension is not considered here. The Example-A
benchmark, which contains a group-delay zero pair, is therefore synthesized in
Section~\ref{sec:results} from its published characteristic polynomials rather
than through this front end.

\subsection{Front-End Results}
For the symmetric dual-band case of Section~\ref{sec:hardware}, the alternation
system reaches a residual below $10^{-28}$ in well under one second. A
more demanding test is the quad-band Example-B specification
of~\cite{shang2012}, which contains twelve prescribed TZs. The outer-band
return loss is fixed at $20$~dB and the inner-band return loss is the freed
condition of \eqref{eq:overconstraint}. From a single cold start,
the solve converges in $6.5$~s to a residual of $7.0\times10^{-28}$. Evaluated
on a $2\times10^{5}$-point frequency grid, the four passbands are equiripple
to within $10^{-4}$~dB.

The optimized inner-band return loss is $30.637$~dB. Although
\cite{shang2012} reports the value as $30$~dB, the published coupling matrix
gives approximately $30.6$~dB when evaluated directly. The recovered
$q_e=1.5327$ agrees with the published value to all four reported decimal
places. The resulting reflection response also agrees closely with the
published design:
\begin{equation}
\max_{\Omega\in[-2,2]}\bigl|\,|S_{11}(\jj\Omega)|
-|S_{11}^{\mathrm{pub}}(\jj\Omega)|\,\bigr|\le 8.4\times10^{-4},
\label{eq:s11cmp}
\end{equation}
on a uniform $2001$-point grid.

\section{Synthesis Results}
\label{sec:results}

\subsection{Benchmark Reproduction}
We first test whether the procedure of Algorithm~\ref{alg:pipeline} reproduces
two published coupling-matrix synthesis benchmarks from~\cite{shang2012}. In
these experiments, the target response is obtained from the published coupling
matrix, so the test isolates the synthesis procedure from the front end of
Section~\ref{sec:frontend}.

Example~A is the tenth-order asymmetric dual-band filter of
Fig.~\ref{fig:topoA}, with different return-loss levels in the two passbands
and a group-delay equalization zero pair. Example~B is a sixteenth-order
quad-band filter with twelve transmission zeros and nine cross-couplings, for
which the response is symmetric and all diagonal entries are constrained to
zero.

Table~\ref{tab:bench} compares the two routes of
Section~\ref{sec:pipeline} on both benchmarks. Every rotation seed is carried
through to a finished design, so that no selection enters the success counts,
and both routes use forward-mode Jacobians, so that the timings compare the
routes rather than two implementations.\footnote{Our original Route-R driver
kept only the best rotation seed and continued from that one, which yields a
design but not a per-seed success rate, and it built its Jacobians in reverse
mode, which for these tall residual vectors costs about twelve times more per
seed with bit-identical residuals. The earlier version of this table
overstated the direct route's advantage on both counts.}

The rotation stage behaves differently at the two filter orders. At $N=10$, it
reaches the prescribed topology from $10$ of $24$ starts, and after zeroing
the residual numerical noise, the resulting designs have errors below
$2.8\times10^{-12}$. At $N=16$, the rotation reaches the topology from none of
the $80$ starts, consistent with the matched benchmark in
Section~\ref{sec:bench16}. Route~R can nevertheless succeed at this order
because its rotation output is used only to initialize the subsequent
hard-masked polish. With this initialization, the full Route~R pipeline
produces a finished design from $62$ of $80$ starts.

Judged by the success of the completed designs, the two routes perform
similarly at the tested orders. Route~D succeeds in $18$ of $24$ starts at
$N=10$ and $12$ of $12$ at $N=16$, compared with $15$ of $24$ and $62$ of
$80$ for Route~R. These differences are not statistically significant at the
available sample sizes (Fisher's exact $p=0.53$ and $p=0.11$), but this does
not establish equivalence; such a claim would require a prespecified
equivalence margin and an appropriate equivalence test. The best runs of both
routes reach comparable errors at the $10^{-15}$ level.

The direct route is also less expensive per successful design, by factors of
$1.6$ at $N=10$ and $1.3$ at $N=16$, when the one-time unconstrained fit
required by Route~R is included and the corresponding cost is excluded from
Route~D. This comparison is against our rotation implementation. As shown in
Section~\ref{sec:bench16}, the published method of~\cite{wang2025} performs
better than our rotation parameterization on its benchmark reconfiguration
problems, but it also fails on the Example-B rotation problem. The practical
choice between the two routes is discussed in Section~\ref{sec:route}.

In both examples, the recovered coupling matrix need not match the published
matrix entry by entry. A prescribed topology does not generally determine a
unique set of coupling values, so the relevant criteria are whether the
synthesized matrix satisfies the prescribed topology and reproduces the target
response. The low computational cost also makes it practical to search for
multiple response-equivalent realizations when a particular coupling
distribution is preferable for implementation.

\begin{table}[htbp]
\caption{Synthesis benchmarks, one CPU core, forward-mode Jacobians
throughout. Route~D: Algorithm~1 (LM on the polynomial residuals, then LM on
the $S$ residuals), i.e.\ variant V3 of Table~\ref{tab:ablation}. Route~R:
unconstrained fit, then the response-preserving rotation of
Section~\ref{sec:rotation}, then a hard-masked polish on the polynomial and
then the $S$ residuals. Example~A: $N{=}10$, 2 passbands, 6 TZs, 25 free
couplings, 28 rotation angles; Example~B: $N{=}16$, 4 passbands, 12 TZs, 24
free couplings, 91 rotation angles. Every seed is carried through to a
finished design; no selection. Success is $\max|\Delta S|<10^{-9}$, as in
Table~\ref{tab:ablation}.}
\label{tab:bench}
\centering
\footnotesize
\setlength{\tabcolsep}{4pt}
\begin{tabular}{lcccc}
\toprule
 & \multicolumn{2}{c}{Example A} & \multicolumn{2}{c}{Example B}\\
\cmidrule(lr){2-3}\cmidrule(lr){4-5}
 & D & R & D & R\\
\midrule
Seeds run             & 24 & 24 & 12 & 80\\
Rotation on topology  & --- & 10/24 & --- & \textbf{0/80}\\
Successes             & 18/24 & 15/24 & 12/12 & 62/80\\
Best $\max|\Delta S|$   & $1.9$e--$15$ & $1.8$e--$15$
                        & $2.5$e--$15$ & $1.9$e--$15$\\
Unconstrained fit     & --- & 1.7 s & --- & 27.1 s\\
Multistart time       & 36.7 s & 47.6 s & 72.6 s & 469 s\\
Time per success      & \textbf{2.0 s} & 3.3 s & \textbf{6.0 s} & 8.0 s\\
\bottomrule
\multicolumn{5}{l}{\scriptsize ``Rotation on topology'': seeds for which the
rotation alone reaches}\\
\multicolumn{5}{l}{\scriptsize the prescribed topology, before the masked
polish. Fisher exact}\\
\multicolumn{5}{l}{\scriptsize on the success counts, D versus R:
$p=0.53$ (A), $p=0.11$ (B).}
\end{tabular}
\end{table}

\subsection{Validation Off the Optimization Grid}
\label{sec:validation}
Because the final phase minimizes the $S$-parameter residual on a
$401$-point grid, that grid alone does not certify the response accuracy. We
therefore validate the three Route-R designs produced by the two Route-R
drivers. These are the least accurate finished designs considered in this
analysis, so the reported errors bound those of the other routes; the Route-D
designs in Table~\ref{tab:bench} are more accurate still.

Each design is reevaluated on an independent $4001$-point grid shifted by half
a grid step and on a $40001$-point grid. We also perform a grid-independent
comparison of the poles, reflection zeros, and transmission zeros of the
synthesized network with those of the reference. The characteristic
polynomials are computed directly from the coupling matrix using the
Faddeev--LeVerrier recursion applied to
$\bm A(s)=s\bm I-(\jj\mM-\bm q)$, yielding $\det\bm A$ and
$\operatorname{adj}\bm A$ and hence $E$, $F$, and $P$ without interpolation.

Table~\ref{tab:validate} reports the results. All three columns correspond to
Route-R runs from Table~\ref{tab:bench}: the two Example-B columns use the same
run before and after the $S$-parameter polish, while the Example-A column uses
the corresponding Route-R result. The Route-D designs in
Table~\ref{tab:bench} reach smaller errors still. The $401$-point evaluation is
therefore not overly favorable: the independent $4001$-point grid and the
$40001$-point grid give values within $10\%$ of it.

The table also illustrates why a small polynomial residual is not sufficient at
high order. Before the $S$-parameter polish, the sixteenth-order design has a
polynomial residual of $6.6\times10^{-11}$, yet its poles are displaced by
$3.7\times10^{-2}$ and its transmission zeros by $1.2\times10^{-1}$, resulting
in $\max|\Delta S|=1.51$. After the polish, the response error drops to the
machine-precision level.

\begin{table}[htbp]
\caption{Validation on grids independent of the optimization grid, and
grid-free root comparison against the reference network. Root errors are
maximum absolute displacements after nearest-neighbour matching.}
\label{tab:validate}
\centering
\footnotesize
\setlength{\tabcolsep}{4pt}
\begin{tabular}{lccc}
\toprule
 & Ex.~B, R & Ex.~B, R & Ex.~A, R\\
 & (after $S$ polish) & (before $S$ polish) & \\
\midrule
$\max|\Delta S|$, 401 pts & $2.1$e--$14$ & $1.510$ & $2.9$e--$12$\\
\quad 4001 pts, offset    & $2.3$e--$14$ & $1.529$ & $2.9$e--$12$\\
\quad 40001 pts           & $2.3$e--$14$ & $1.529$ & $2.9$e--$12$\\
Pole error                & $7.4$e--$12$ & $3.7$e--$2$ & $1.3$e--$14$\\
Reflection-zero error     & $1.0$e--$10$ & $7.7$e--$2$ & $9.3$e--$13$\\
Transmission-zero error   & $7.9$e--$10$ & $1.2$e--$1$ & $3.0$e--$13$\\
\bottomrule
\end{tabular}
\end{table}

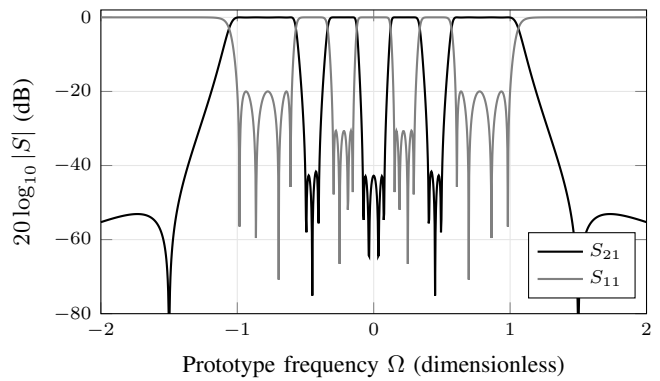
\begin{figure}[!t]
\centering
\begin{tikzpicture}
\begin{axis}[width=8.8cm, height=5.6cm,
  xlabel={Prototype frequency $\Omega$ (dimensionless)},
  ylabel={$20\log_{10}|S|$ (dB)},
  xmin=-2, xmax=2, ymin=-80, ymax=2,
  legend style={font=\scriptsize, at={(0.98,0.05)}, anchor=south east},
  grid=both, grid style={gray!20}, tick label style={font=\scriptsize},
  label style={font=\small}]
\addplot[black, thick] table[x=omega, y=s21db] {data/exampleB.dat};
\addplot[black!50, thick] table[x=omega, y=s11db] {data/exampleB.dat};
\legend{$S_{21}$, $S_{11}$}
\end{axis}
\end{tikzpicture}
\caption{Synthesized response for the sixteenth-order quad-band benchmark
(Example~B), using the most accurate of the $12$ Route-D seeds in
Table~\ref{tab:bench}. The four passbands, twelve transmission zeros, and
return-loss levels of $20$~dB in the outer bands and $30.6$~dB in the inner
bands are reproduced with
$\max|\Delta S|=2.5\times10^{-15}$ on the metric grid and
$5.1\times10^{-15}$ on an independent $40001$-point grid. The $30.6$~dB
inner-band level is the optimized value of the freed condition in
\eqref{eq:overconstraint}. The response is plotted on $1201$ points.}
\label{fig:respB}
\end{figure}

\subsection{Specification-to-Matrix Validation on a Fabricated-Filter
Specification}
\label{sec:hardware}

The previous benchmarks begin from known target polynomials. We next test the
complete chain of Fig.~\ref{fig:pipeline}, beginning instead from the
specification of the fabricated eighth-order dual-band $X$-band waveguide
filter of~\cite{shang2012}. The prototype specification contains passbands at
approximately $\pm[0.46,1.0]$, transmission zeros at $\pm\jj0.2$, a $20$-dB
return-loss requirement, and the folded topology with cross-couplings $1$--$4$
and $5$--$8$.

Before synthesizing a new matrix, we first reverse-engineer the prototype
response represented by the published rounded coupling matrix. The coefficients
reported to four decimal places need not reproduce the nominal specification
exactly. Throughout this subsection, ``reference response'' therefore denotes
the response obtained from the published rounded coefficients, rather than an
idealized equiripple prototype. The recovered band edges and transmission-zero
locations are then used as the specification for the end-to-end test.

Starting only from the recovered specification, the full synthesis chain
completes in $85.9$~s and produces the coupling values listed in
Table~\ref{tab:coup}. The resulting matrix reproduces the idealized target
defined by that specification, namely the exactly equiripple prototype, to
numerical precision, with an in-band return loss of $19.999$~dB on a
$2\times10^{5}$-point grid.

Most notably, the synthesis also converges to a solution in which one of the two published
cross-couplings vanishes, with $|m_{58}|=5.9\times10^{-14}$. The specification
driven run alone does not establish whether this reduced topology can
reproduce the response of the published matrix, because the rounded
coefficients do not realize the idealized specification exactly. The published
matrix gives an in-band return loss of $19.971$~dB rather than $20.000$~dB,
and its response differs from the idealized target by up to $0.256$ in
$|S_{11}|$ over $\Omega\in[-2,2]$, or $0.073$ within the passbands.

One convention first: the synthesis drives $m_{14}$, not $m_{58}$, to zero.
Since $q_{e1}=q_{eN}$, reversing the resonator order $i\mapsto N{+}1{-}i$ maps
the two orientations onto each other with the response unchanged
($\max|\Delta S|=4.4\times10^{-14}$ on $2001$ points), and we report the
reversed one so that the vanishing entry matches the published indexing.

We test reachability directly in both directions, using the exact
characteristic polynomials of each target and enforcing
$m_{58}\equiv0$ from the first iteration, with $40$ random starts for each
target.

\emph{Result.} For the idealized equiripple target, the folded topology with
one cross-coupling is sufficient: the constrained fit reaches
$\max|\Delta S|=5.0\times10^{-14}$, compared with
$1.6\times10^{-15}$ for the unconstrained topology. For the response of the
published rounded matrix, however, it is not. The unconstrained topology
reproduces that response from the first seed with
$\max|\Delta S|=2.4\times10^{-14}$, whereas enforcing $m_{58}\equiv0$ gives a
best residual of $2.7$ and $\max|\Delta S|=0.416$ over the $40$ starts.

The supported claim is therefore narrower than one about the published
coupling matrix itself. The \emph{specification} of the fabricated filter,
with band edges $\pm[0.46,1.0]$, transmission zeros at $\pm\jj0.2$, and
$20$-dB return loss, admits a folded realization with one fewer
cross-coupling. For the response actually produced by the published rounded
coefficients, no such realization was found: none of the $40$ constrained
starts reached $\max|\Delta S|<0.416$, whereas the unconstrained topology
reproduced the same response to $2.4\times10^{-14}$ from the first seed. This
is strong numerical evidence, not a proof of non-existence, because the search
is local and finitely sampled. We therefore do not claim that the sparser
topology is incapable of realizing that response. The contrast between the
two cases is nevertheless useful for design: the reduction is available for
the response specified by the designer, while the $0.029$-dB shift introduced
by rounding appears sufficient to remove it for the published realization.

This is a prototype-level result, not a claim about the final electromagnetic
structure. The sparse realization is not front-to-back symmetric, unlike the
published waveguide design, and it remains to be established whether its
coupling signs and magnitudes can be implemented conveniently with the same
physical iris configuration.

\begin{table}[htbp]
\caption{Fabricated dual-band filter: published versus synthesized. One
\emph{signed} realization of this work's design is given, at the precision
needed to reproduce the response; the coupling graph contains the single cycle
$1$--$2$--$3$--$4$--$1$, so all other sign patterns are reachable from this one
by a diagonal $\pm1$ similarity, and the cycle product is the one
similarity-invariant sign quantity. The resonator index order is fixed so that
the vanishing cross-coupling is $m_{58}$ rather than the $m_{14}$ the
synthesis produces; the two orientations are related by reversing the resonator
order and are response-equivalent because $q_{e1}=q_{eN}$ (see text).}
\label{tab:coup}
\centering
\footnotesize
\begin{tabular}{lcc}
\toprule
 & Published~\cite{shang2012} & This work\\
\midrule
$q_e$    & 1.7278 & 1.746500\\
$m_{12}$ & $\phantom{-}0.6452$ & $-0.624287$\\
$m_{23}$ & $\phantom{-}0.0476$ & $\phantom{-}0.051887$\\
$m_{34}$ & $\phantom{-}0.6623$ & $-0.618882$\\
$m_{45}$ & $\phantom{-}0.3786$ & $-0.388377$\\
$m_{56}$ & $\phantom{-}0.6623$ & $-0.707564$\\
$m_{67}$ & $\phantom{-}0.0476$ & $\phantom{-}0.443827$\\
$m_{78}$ & $\phantom{-}0.6452$ & $\phantom{-}0.823692$\\
$m_{14}$ & $-0.5389$ & $-0.537339$\\
$m_{58}$ & $-0.5389$ & $\bm{5.9\times10^{-14}}$\\
\midrule
$m_{12}m_{23}m_{34}m_{41}$ & $-1.0961\times10^{-2}$ & $-1.0772\times10^{-2}$\\
\bottomrule
\end{tabular}
\end{table}

\begin{figure}[!t]
\centering
\begin{tikzpicture}
\begin{axis}[width=8.8cm, height=5.6cm,
  xlabel={Prototype frequency $\Omega$ (dimensionless)},
  ylabel={$20\log_{10}|S|$ (dB)},
  xmin=-2, xmax=2, ymin=-70, ymax=2,
  legend style={font=\scriptsize, at={(0.98,0.05)}, anchor=south east},
  grid=both, grid style={gray!20}, tick label style={font=\scriptsize},
  label style={font=\small}]
\addplot[black, thick] table[x=omega, y=s21db] {data/dualband.dat};
\addplot[black!50, thick] table[x=omega, y=s11db] {data/dualband.dat};
\addplot[black!50, thick, densely dotted] table[x=omega, y=s11ydb]
        {data/dualband.dat};
\addplot[gray, dashed, domain=-2:2] {-20};
\legend{$S_{21}$, $S_{11}$ nominal, $S_{11}$ softplus-opt.}
\end{axis}
\end{tikzpicture}
\caption{Eighth-order dual-band filter synthesized directly from the
specification ($85.9$~s, single core). Solid: nominal, exactly equiripple
design (realized with $m_{58}=0$). Dotted: the additive-model
\emph{mean-softplus} yield-centered design of Section~\ref{sec:yield}, which
trades $0.58$~dB of nominal margin for robustness (the LSE variant trades
$1.02$~dB). Both figures are for the first training draw; the five-draw means
are $0.68\pm0.17$ and $1.11\pm0.10$~dB.}
\label{fig:respDB}
\end{figure}
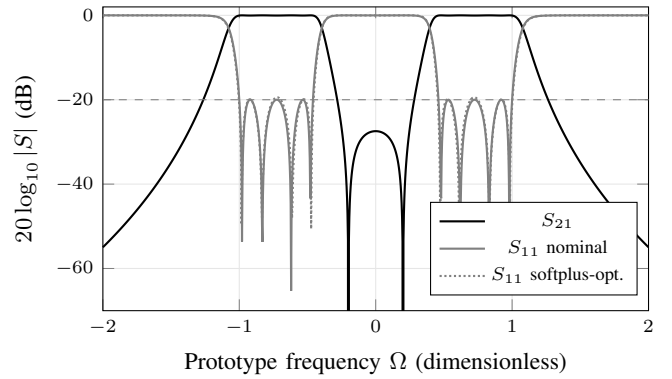

\subsection{Topology Stress Tests and Infeasibility Behavior}
\label{sec:stress}

The two literature benchmarks demonstrate accuracy on established designs but
do not show how the method behaves across different coupling graphs. We
therefore add a stress test over four representative structures, with two
independently generated target matrices each.

For each target, mainline couplings are sampled from $\mathcal U(0.4,0.9)$,
cross-couplings from $\mathcal U(0.2,0.6)$ with random signs, and, for
asymmetric cases, self-couplings from $\mathcal U(-0.3,0.3)$; $q_e=1.2$. The
target polynomials are computed from the resulting matrix, so the ground-truth
topology and response are known exactly. The structures are cascaded triplets,
the folded canonical form, cascaded quadruplets and a wheel topology. The
masks and random seeds are included in the released code.

All eight feasible cases converge, with $\max|\Delta S|\le7\times10^{-14}$ and
run times of $7$--$10$~s each (Table~\ref{tab:stress}).

We also deliberately test two topologies that cannot represent the chosen
Example-A target response: one whose support admits no finite transmission
zero, and one that admits a single triplet. Here the \emph{rotation} formulation of Section~\ref{sec:rotation} is
informative even though it is not the recommended synthesis route, because its
residual measures directly how far the response-preserving orbit of $\mM_0$
comes from the prescribed support. In both cases the rotation residual terminates at a large
nonzero floor---$8.25\times10^{-2}$ and $5.11\times10^{-2}$, with minimum and
median coinciding to three digits over $40$ random seeds, whereas every
feasible case in Table~\ref{tab:stress} reaches $10^{-28}$ or below, and the
largest rotation residual for any feasible topology anywhere in this paper is
$3.0\times10^{-4}$ (Example~B).

We describe this as a \emph{heuristic indicator} of topological infeasibility,
not a certificate and not a graded measure. A nonzero floor can equally result
from an insufficient number of starts, a deficient parameterization or local
optimizer failure, and two deliberately constructed masks are far too few to
establish that its magnitude grades the degree of topology deficit. The
observation that the floor decreases when the mask is enlarged by one
transmission-zero-producing coupling is suggestive and no more. Useful
thresholds are also problem-dependent: the feasible sixteenth-order benchmark
itself stops at $3\times10^{-4}$. A formal characterization of reachability,
including the discrete sign-diagonal and permutation equivalences and the
source--load-coupled topologies that require the $N{+}2$ coupling-matrix
formulation~\cite{cameron2003}, which the $N\times N$ model of
Section~\ref{sec:model} does not cover, remains open; \cite{lee2025} is the most complete
treatment we are aware of.

\begin{table}[!t]
\caption{Topology stress tests (two random targets per topology; success
threshold $\max|\Delta S|<10^{-9}$).}
\label{tab:stress}
\centering
\footnotesize
\begin{tabular}{lccc}
\toprule
Topology & $N$ & $\max|\Delta S|$ & Time\\
\midrule
Cascaded triplets (asym.) & 6 & $\le6.0\times10^{-14}$ & 7--10 s\\
Folded canonical & 8 & $\le1.0\times10^{-14}$ & 8--9 s\\
Cascaded quadruplets & 8 & $\le1.3\times10^{-15}$ & 7 s\\
Wheel & 6 & $<10^{-15}$ & 10 s\\
\midrule
Infeasible: mainline-only$^{\dagger}$ & 10 &
\multicolumn{2}{c}{rotation floor $8.25\times10^{-2}$}\\
Infeasible: single triplet$^{\dagger}$ & 10 &
\multicolumn{2}{c}{rotation floor $5.11\times10^{-2}$}\\
\bottomrule
\multicolumn{4}{l}{\scriptsize $^{\dagger}$Example-A targets (six finite TZs);
the masks support zero and one TZ.}
\end{tabular}
\end{table}

\section{Yield-Aware Synthesis}
\label{sec:yield}

The differentiable formulation extends beyond nominal synthesis to
tolerance-aware design. Because the $S$-parameter calculation remains
differentiable after perturbing the couplings, gradients can be propagated
through a Monte Carlo (MC) tolerance simulation, so that the nominal coupling
matrix itself can be adjusted to improve the predicted yield. Throughout this
section ``yield'' means \emph{model-predicted} yield under the stated
tolerance model; no measurement or electromagnetic model enters. Yield-driven
design and design centering are of course long established, going back for
arbitrary statistical distributions to Abdel-Malek and
Bandler~\cite{abdelmalek1980}; modern microwave-filter
treatments include polynomial-chaos yield-constrained
optimization~\cite{zhang2021yield} and the review
of~\cite{zhang2025robust}. What AD contributes is not a new yield formulation
but the ability to differentiate an arbitrary tolerance model and acceptance
surrogate without deriving new sensitivity expressions, which is what made the
model and surrogate comparisons below cheap enough to run with paired
controls.

\subsection{Tolerance Models}
Let $\xi_{ij}\sim\mathcal N(0,1)$ be independent standard normal variates, one
per \emph{undirected} coupling allowed by the topology, and set
$\xi_{ji}=\xi_{ij}$ so that the perturbed network stays reciprocal. The
eight-resonator topology used here has nine such couplings and therefore nine
independent random variables per sample. In the additive model the realized
matrix is $\mM+\bm\Delta$ with
\begin{equation}
\Delta m_{ij}=\sigma\,\xi_{ij},\qquad \sigma=0.005,
\label{eq:addmodel}
\end{equation}
which is about $1\%$ for a mid-sized coupling ($|m|\approx0.5$) but a much
larger relative error for a small coupling such as $|m_{23}|\approx0.05$; we
therefore call it an additive coupling-error model rather than a uniform
relative tolerance. The multiplicative model instead uses
\begin{equation}
\Delta m_{ij}=\rho\,m_{ij}\,\xi_{ij},\qquad \rho=0.01,
\label{eq:multmodel}
\end{equation}
so that a coupling that is exactly zero receives no perturbation. Only
coupling errors are modeled; resonator-frequency detuning, external-$Q$
error, dimensional sensitivity and correlations between errors are not.

\subsection{Acceptance Criterion and Objective}
The optimization topology is the mainline couplings together with $m_{14}$ and
$m_{58}$, with $m_{58}$ initialized at zero. A realization is acceptable when
the in-band return loss is at least $17$~dB; the nominal design is $20$-dB
equiripple, leaving a $3$-dB margin. To keep the acceptance quantity
differentiable everywhere, including at an exact reflection zero where
$20\log_{10}|S_{11}|$ is singular, we use the stabilized magnitude
\begin{equation}
L(\Omega;\mM)=10\log_{10}\!\bigl(|S_{11}(\jj\Omega;\mM)|^{2}+\delta\bigr),
\qquad \delta=10^{-12}.
\label{eq:stab}
\end{equation}
Equation \eqref{eq:stab} is the quantity that is differentiated and also the
one used for evaluation. The acceptance test is a maximum over the passbands,
where $|S_{11}|$ is close to $-20$~dB, so the stabilization is inactive in
practice: an earlier version of these experiments used
$20\log_{10}(|S_{11}|+10^{-300})$ and every yield agrees with the stabilized
result to within $0.005$ percentage points.

Directly optimizing the fraction of MC samples that pass would give a
discontinuous objective, so we use a smooth surrogate. With $K=256$
perturbation samples $\bm\Delta_k$ drawn once and held fixed (common random
numbers) and $\mathcal B$ the union of the passbands, the two surrogates
compared here are
\begin{align}
\mathcal{L}_{\mathrm{SP}}(\mM)&=\frac1K\sum_{k=1}^{K}
\ \underset{\Omega\in\mathcal B}{\mathrm{mean}}\ \
\sigma_{+}\!\Bigl(\tfrac{1}{\tau}\bigl[L(\Omega;\mM{+}\bm\Delta_k)-T\bigr]\Bigr),
\label{eq:yieldsp}\\[2pt]
\mathcal{L}_{\mathrm{LSE}}(\mM)&=\frac1K\sum_{k=1}^{K}
\sigma_{+}\!\Bigl(\tfrac{1}{\tau}\bigl[
\Lambda_k(\mM)-T\bigr]\Bigr),
\label{eq:yieldlse}\\[2pt]
\Lambda_k(\mM)&=\tau_\ell\log\!\!\sum_{\Omega_i\in\mathcal B}\!
\exp\!\bigl(L(\Omega_i;\mM{+}\bm\Delta_k)/\tau_\ell\bigr),
\label{eq:lse}
\end{align}
where $\sigma_{+}$ is the softplus function, $T=-17$~dB, $\tau=0.5$~dB and
$\tau_\ell=0.15$~dB. Equation \eqref{eq:lse} is a log-sum-exp smooth maximum,
which upper-bounds $\max_i L(\Omega_i)$ by at most $\tau_\ell\log|\mathcal B|$
and therefore targets the worst in-band value that actually determines
acceptance, whereas \eqref{eq:yieldsp} penalizes the response throughout the
passbands. Because the same perturbations are reused at every step, both
surrogates are deterministic differentiable functions of the nominal
couplings: \texttt{jax.grad} propagates derivatives through all $K$ perturbed
$S$-parameter evaluations and \texttt{jax.vmap} vectorizes them. Four hundred
Adam steps recenter the design in $22$--$24$~s on one core.

\subsection{Choice of $\sigma$}
The value $\sigma=0.005$ is selected by a short sweep rather than fixed
a priori. The released script evaluates
$\sigma\in\{0.002,0.003,0.005,0.008,0.012,0.018,0.025\}$ and reports the
nominal yield at each. We selected the first tested value, in ascending order,
for which that yield lies between $15\%$ and $75\%$, keeping the comparison in
a range where yield can change in either direction; the first two values give
$98.5\%$ and $87.3\%$, so the criterion selects $\sigma=0.005$, which is the
value hard-coded in the yield scripts. Table~\ref{tab:sigma} reports the whole sweep.
The ordering of the two nominal designs is unchanged at all seven tested
values of $\sigma$ under the additive model, so the conclusion in
Section~\ref{sec:yieldresults} does not depend on the stopping point of the
sweep.

\begin{table}[htbp]
\caption{Model-predicted yield versus the additive tolerance $\sigma$
($4096$-sample MC, common random numbers, $40$ in-band samples per passband).
The ordering of the two nominal designs is unchanged across the range, which
is the point of the table; absolute values on the converged $400$-sample grid
are $0.4$--$0.6$ points lower (Table~\ref{tab:yield}).}
\label{tab:sigma}
\centering
\footnotesize
\setlength{\tabcolsep}{4pt}
\begin{tabular}{lccccccc}
\toprule
$\sigma$ & 0.002 & 0.003 & \textbf{0.005} & 0.008 & 0.012 & 0.018 & 0.025\\
\midrule
This work  & 98.5 & 87.3 & \textbf{51.3} & 17.7 & 4.2 & 0.7 & 0.2\\
Published  & 96.0 & 82.4 & \textbf{46.8} & 15.7 & 3.5 & 0.5 & 0.2\\
\bottomrule
\end{tabular}
\end{table}

\subsection{Results and Controls}
\label{sec:yieldresults}
Table~\ref{tab:yield} reports hard Monte Carlo yields on $4096$ fresh
perturbation draws. The two sources of uncertainty are reported separately.
The \emph{training-draw} spread is the sample standard deviation over five
independent draws of the $K=256$ common-random-number training set, with each
design retrained from scratch. The \emph{finite-MC} uncertainty is obtained
from paired evaluations: both designs are evaluated on the same $4096$ draws,
the sample-by-sample yield differences are formed, and a paired bootstrap is
used to construct the interval. All comparisons reported below use paired
samples.

The acceptance test was checked for grid convergence using $40$, $200$, and
$400$ in-band samples per passband. Refinement from $200$ to $400$ samples
reduces every yield by $0.4$--$0.6$ percentage points and changes no ranking.
The values reported throughout use the converged $400$-sample evaluation; all
three grid resolutions are included in the archive.

\begin{table}[htbp]
\caption{Model-predicted yield (spec: in-band RL $\ge17$~dB; $4096$-sample MC,
$400$ in-band samples per passband). Centered rows are mean\,$\pm$\,sample SD
over five independent training draws, each trained under the tolerance model
of its own column, so the two columns of a centered row contain
\emph{different designs}. LSE$_{\mathrm c}$ is the maximum-offset-corrected
(log-mean-exp) control of \eqref{eq:lsec}. The published-matrix entry is not the baseline for
centering; the baseline is the nominal design of Section~\ref{sec:hardware}.}
\label{tab:yield}
\centering
\footnotesize
\begin{tabular}{lcc}
\toprule
Design & Add.\ $\sigma{=}0.005$ & Mult.\ $\rho{=}1\%$\\
\midrule
Published (GA+SQP)~\cite{shang2012} & 46.3\% & 46.9\%\\
This work, equiripple nominal & 50.9\% & 44.7\%\\
\ \ centered, mean-softplus & $65.9\pm0.3$\% & $56.3\pm0.6$\%\\
\ \ centered, LSE$_{\mathrm c}$ & $66.5\pm0.9$\% & $59.2\pm0.6$\%\\
\ \ centered, LSE & $\bm{67.0\pm0.5}$\% & $\bm{59.9\pm0.3}$\%\\
\bottomrule
\end{tabular}
\end{table}

\begin{figure}[H]
\centering
\begin{tikzpicture}
\begin{axis}[width=8.8cm, height=5.6cm,
  xlabel={Worst-case in-band $10\log_{10}(|S_{11}|^2+\delta)$ (dB)},
  ylabel={Cumulative probability},
  xmin=-20, xmax=-12, ymin=0, ymax=1,
  legend style={font=\scriptsize, at={(0.02,0.98)}, anchor=north west},
  grid=both, grid style={gray!20}, tick label style={font=\scriptsize},
  label style={font=\small}]
\addplot[black!50, thick] table[x=wnom, y=cdf] {data/yield_cdf.dat};
\addplot[black, thick] table[x=wopt, y=cdf] {data/yield_cdf.dat};
\addplot[black, thick, densely dashed] table[x=wlse, y=cdf]
        {data/yield_cdf.dat};
\addplot[gray, dashed] coordinates {(-17,0) (-17,1)};
\legend{nominal, mean-softplus opt., LSE opt., spec}
\end{axis}
\end{tikzpicture}
\caption{Distribution of the worst-case in-band return loss under the additive
coupling-error model ($\sigma=0.005$; $4096$ Monte Carlo samples, common
random numbers, first training draw, $40$ in-band samples per passband). Both
centered designs shift the distribution left; the fraction left of the
$-17$-dB spec line is the yield ($51.3\%$, $66.6\%$ and $67.2\%$ on this
grid).}
\label{fig:cdf}
\end{figure}
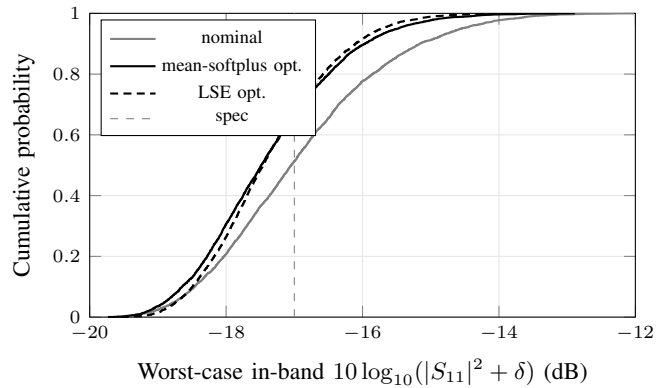

\emph{Effect of design centering.} Starting from the $50.9\%$ nominal design,
the LSE objective increases the hard Monte Carlo yield to $67.0\pm0.5\%$ on the
converged $400$-sample acceptance grid, a paired improvement of $16.1$
percentage points with a $95\%$ bootstrap CI of $[14.6,17.6]$. The interval is
a percentile interval for exactly that quantity: the sample-by-sample mean
paired difference across the five centered designs, resampling the $4096$
common perturbation draws jointly. Under the multiplicative model the paired
gain is $15.1$ percentage points, CI $[13.8,16.4]$.

The improvement does not come from increasing the nominal worst-case return
loss. Averaged over the five training draws the centered designs give up
$0.68\pm0.17$~dB for mean-softplus and $1.11\pm0.10$~dB for LSE (the first
draw, plotted in Fig.~\ref{fig:respDB}, gives up $0.58$ and $1.02$~dB), while
shifting the full distribution in
Fig.~\ref{fig:cdf} by approximately $0.4$--$0.8$~dB. The two effects should be
distinguished from the $46.3\%$ yield of the published matrix, which reflects
the coefficient values present in the published four-decimal matrix. We
therefore
report the two changes separately: $50.9\%\rightarrow67.0\%$ from design
centering, and $46.3\%\rightarrow50.9\%$ from resynthesis. These effects are
not combined.

\emph{Centering reactivates the dormant cross-coupling.} In all six
model/surrogate combinations and all five training draws, centering moves
$m_{58}$ away from zero, to $|m_{58}|$ between $0.0026$ and $0.0211$. The
centered designs therefore do \emph{not} retain the one-cross-coupling
realization of Section~\ref{sec:hardware}; the sparse design and the centered
design are alternatives, not a single design.

The optimization variables span every coupling of the \emph{folded} topology,
$m_{58}$ included, so centering is free to move it; the two error models then
differ on a nominally zero entry. Additively $m_{58}\mapsto\sigma\xi_{58}$, so
even an untouched zero acquires scatter; multiplicatively
$m_{58}\mapsto m_{58}(1+\rho\xi_{58})$ leaves an exact zero exactly zero until
centering moves it. Modelling a physically \emph{absent} edge is therefore not
a matter of constraining the perturbation: it requires removing $m_{58}$ from
the optimization variables altogether. The yields quoted here are those of the
folded topology evaluated at $m_{58}=0$, not of a structurally sparser filter.

\emph{The surrogate effect is real under multiplicative errors and not
established under additive ones.} The log-sum-exp smooth maximum
\eqref{eq:lse} overestimates the true in-band maximum by up to
$\tau_\ell\log|\mathcal B|=0.66$~dB for the $80$ in-band samples used here, so
part of any advantage it shows could be mere conservatism---it is effectively
optimizing against a stricter threshold. We therefore also ran a
maximum-offset-corrected control
\begin{equation}
\Lambda^{\mathrm c}_k(\mM)=\Lambda_k(\mM)-\tau_\ell\log|\mathcal B|,
\label{eq:lsec}
\end{equation}
which subtracts the largest offset the smooth maximum can carry, converting
log-sum-exp into log-mean-exp. This does not cancel the bias pointwise: the
correction is exact only when all $L(\Omega_i)$ are equal, and it
\emph{under}-estimates the true maximum whenever one frequency dominates.
The two forms therefore bracket $\max_i L(\Omega_i)$ from above and below, and
the useful test is whether the surrogate effect persists across the bracket.
Under multiplicative errors LSE beats mean-softplus by $+3.59$ points
($95\%$ CI $[+2.66,+4.51]$), and the less conservative control still beats it
by $+2.93$ points (CI $[+1.95,+3.93]$): the effect persists at both ends of the
bracket, so it reflects targeting the worst in-band value rather than the
offset. Under additive errors LSE beats mean-softplus by only $+1.17$ points
(CI $[+0.25,+2.08]$) and the corrected control by $+0.60$ points
(CI $[-0.34,+1.54]$, which includes zero). We therefore claim a surrogate effect under the multiplicative model
only. The offset itself is worth $+0.58$ and $+0.65$ points in the two models.

\emph{The preferred nominal design depends on the tolerance model.} This
comparison involves no optimizer and no training draw, and both designs are
evaluated on the same $4096$ perturbations, so it is tested by a paired
bootstrap. Under additive errors this work's sparse, front--back asymmetric
design beats the published one by $+4.64$ points ($95\%$ CI $[+2.64,+6.62]$);
under multiplicative errors it \emph{loses} by $2.20$ points (CI
$[-3.91,-0.54]$). Both intervals exclude zero, so the reversal is real for
these two designs under these two models. The mechanism is that the sparse
realization concentrates the response into fewer, larger couplings, which
receive correspondingly larger absolute perturbations when the error scales
with coupling magnitude.

That said, the two designs differ in $q_e$, in symmetry, in coupling
distribution and in nominal ripple accuracy, so this compares two specific
realizations rather than isolating a sparsity effect. Establishing the latter
would require synthesizing exact nominal designs for the symmetric
two-cross-coupling, sparse one-cross-coupling and redundant topologies and
cross-evaluating them under identical margins, grids and sample sets; we have
not done so.

\section{Practical Guidelines}
\label{sec:practice}

The results support the following practical procedure for differentiable
coupling-matrix synthesis.

\emph{Objective.} Fit the characteristic polynomials first and the
$S$-parameters second, as shown in Table~\ref{tab:ablation}. Use the relative
residuals in \eqref{eq:relres} with a sample count on the order of the
polynomial degree; in the one denser configuration tested, increasing the
sample count substantially did not help and degraded conditioning
(Table~\ref{tab:rings}). The choice of sampling radius has little effect over
the range tested.

\emph{Optimizer.} Use Levenberg--Marquardt in both phases, with forward-mode
Jacobians when the residual vector is tall, and a budget of at most $300$
iterations per phase. The stochastic-gradient phase did not improve success in
these tests and costs roughly double the wall time, so we omit it
(Table~\ref{tab:ablation}); we do not claim it is equivalent. If Adam is used, its step size should be tuned
carefully; the learning-rate sweep in Table~\ref{tab:lrsweep} shows a much
stronger dependence on step size than the usual default would suggest.

\emph{Initialization.} Use $12$--$24$ independent starts with
$m_{ij}\sim\mathcal U(-0.7,0.7)$ on the allowed entries. The starts can be
evaluated independently and vectorized. Random multistart is not the only
remedy for local minima in coupling-matrix fitting: homotopy continuation has
been combined with Levenberg--Marquardt for coupling-matrix
diagnosis~\cite{zhao2024}, and would be the natural alternative to compare
against. We have not done so, and the choice of objective studied here is
independent of that choice of globalization strategy.

\emph{Diagnosing failure.} Distinguish three cases. If several starts reach a
small polynomial residual but retain a large $\max|\Delta S|$, the polynomial
stage has succeeded and the $S$-parameter phase is still required. This is the
typical behavior at high order (Table~\ref{tab:validate}). If all starts
stall at a similar polynomial residual, the topology may be poorly matched to
the target response. The rotation residual in Section~\ref{sec:rotation} can
serve as a heuristic check, but the orbit ambiguity discussed in
Section~\ref{sec:whyfail} must be taken into account. Finally, if multiple
starts succeed but produce coupling values that are inconvenient for the
intended realization, the topology may admit multiple response-equivalent
matrices. Example~B has a three-parameter family of such solutions
(Section~\ref{sec:redundant}), so additional starts provide a practical way to
search for a more suitable realization.

\emph{Choosing the topology.} The method does not select the topology; the
topology is an input. It instead provides a relatively inexpensive way to
evaluate candidate topologies against the same response and compare their
resulting coupling distributions and, when tolerance data are available,
their predicted yields (Section~\ref{sec:yield}).

\emph{From prototype to hardware.} The results in this paper are limited to
normalized low-pass prototypes. Converting a coupling matrix into physical
dimensions and verifying the resulting structure electromagnetically are
outside the scope of this work. In particular, the $m_{58}$ result in
Section~\ref{sec:hardware} concerns prototype reachability and does not
establish that the corresponding sparse matrix can be realized with the same
waveguide iris configuration.

\section{Discussion and Limitations}
\label{sec:discussion}

\emph{The procedure is staged rather than end-to-end differentiable.}
Each continuous optimization in Fig.~\ref{fig:pipeline} uses
automatic-differentiation Jacobians, which is the differentiability claim made
here. The complete procedure is not end-to-end differentiable: the phases are
separate solves, no derivative is propagated through the solution map of an LM
problem, spectral factorization and Hurwitz-factor selection occur outside the
AD loop, and the centering procedure of Section~\ref{sec:yield} differentiates
only the final forward model, not the synthesis that produced its starting
point. A fully differentiable formulation could instead propagate gradients
through the LM stationarity conditions, for example by implicit
differentiation, allowing a yield or manufacturability objective to influence
the target specification. This is left for future work.

\emph{Multi-start remains necessary.} The direct procedure succeeds from
$18$ of $24$ seeds for Example~A and from all $12$ seeds for Example~B. The
seeds are independent, and the per-seed computations have the same structure,
so the multi-start solve can be vectorized over the seed axis with
\texttt{jax.vmap}. The resulting LM normal equations are batched solves of
size $25\times25$ or $24\times24$. A single successful design currently
requires $2$--$6$~s on one CPU core, so the main benefit of GPU execution would
be concurrent evaluation of many seeds or larger topology and tolerance
sweeps rather than lower latency for one solve. GPU performance was not
measured here.

\emph{Conditioning is worse at sixteenth order, but it does not distinguish the
objectives.} Table~\ref{tab:kappa} shows that the effective condition number
of the polynomial objective is about an order of magnitude larger for
Example~B than for Example~A, consistent with the attainable polynomial
residual increasing from $10^{-28}$ to $10^{-10}$. Table~\ref{tab:validate}
also shows that the latter residual constrains the response only weakly at high
order. The $S$-parameter Jacobian is nevertheless better conditioned at both
orders, so local conditioning does not explain the difference in optimization
success; the residual saturation discussed in \eqref{eq:bound} provides a
more consistent explanation. This comparison rests on the two benchmarks; the
eight synthetic stress cases of Section~\ref{sec:stress} probe convergence and
rotation-residual floors rather than the objective comparison, and are not
evidence for it. Whether the same balance holds
at substantially higher order, or for lossy and frequency-dependent models,
has not been tested. The evidence for saturation consists of the response
bound and scale-sweep experiments rather than a direct measurement of LM
dynamics. A stronger test would record predicted reduction, rejected steps,
and accepted step lengths throughout the iterations.

\emph{The rotation route is not recommended here, and the implementation used
is not the strongest published formulation.} Section~\ref{sec:bench16} shows
that the method of~\cite{wang2025} outperforms the matrix-exponential
parameterization used here on the reference benchmarks, with a larger
difference as the number of internal resonators increases. The measurements
are consistent with the single global parameterization accumulating large
$\|\bm K\|_F$ and poorer Jacobian conditioning, whereas the retraction-based
method re-centers its local coordinates during the optimization. We have not
benchmarked against the other related formulations in
\cite{lee2025,wu2024,zeng2024,wu2025householder,qian2025}.
Several target the larger-$N$ regime in which the present parameterization is
weakest. A broader comparison, particularly on the sixteenth-order problem
that defeats both implementations tested here, remains open.

\emph{Scope of the yield study.} The yield analysis considers one filter, two
tolerance models, three acceptance surrogates, and coupling errors only. The
tolerance scale $\sigma$ was selected to place the nominal yield in a useful
mid-range rather than derived from a specific fabrication process. We also do
not include a controlled three-topology comparison designed to isolate the
effect of sparsity. The reversal reported in Section~\ref{sec:yieldresults}
therefore applies to the two realizations studied here rather than establishing
a general sparsity effect.

\emph{Extensions require changes to the forward model rather than new
sensitivity derivations.} The formulation uses a differentiable numerical
forward model, so extensions such as finite resonator $Q$, frequency-dependent
couplings, resonator-frequency tolerances, or coupling to a differentiable
electromagnetic surrogate can be incorporated by modifying that model and its
inputs. The tolerance study in Section~\ref{sec:yield} illustrates this
modularity: changing the tolerance model or acceptance surrogate does not
require new analytical derivative expressions.

\section{Conclusion}

This work examined what automatic differentiation does and does not contribute
to coupling-matrix synthesis of cross-coupled resonator filters. Automatic
differentiation removes the need to derive and maintain problem-specific
sensitivity expressions, making it straightforward to change objectives,
tolerance models, and acceptance criteria. It does not by itself make the
synthesis problem easy. A controlled ablation with matched initializations,
optimizer, and final refinement shows that the objective used in the first
phase is the dominant factor in optimization success: matching the
frequency-sampled response succeeds from $0$ of $24$ starts at tenth order and
$0$ of $12$ at sixteenth order, whereas matching the characteristic
polynomials succeeds from $18$ of $24$ and $12$ of $12$, respectively. The
successful runs reach $\max|\Delta S|\approx2\times10^{-15}$ and require only
a few seconds per design on one CPU core. Adding a stochastic-gradient phase
produced no detectable improvement at these sample sizes and roughly doubled
the wall time, so it is omitted.

The Jacobian analysis rules out local conditioning as the explanation for the
difference between the two objectives. After accounting for the exact rank
deficiency of the sixteenth-order problem, the $S$-parameter Jacobian is better
conditioned at both orders. The rank deficiency is itself significant:
Example~B has a three-parameter family of response-equivalent coupling
matrices. The remaining evidence is consistent with saturation of the
$S$-parameter residual. Because $|S_{11}|,|S_{21}|\le1$, the response residual
has a bounded dynamic range and changes by only about a factor of $1.4$ over
a region in which the polynomial residual changes by five to seven orders of
magnitude. The resulting Gauss--Newton model provides much less variation far
from the solution. We present this as evidence for the mechanism rather than
as a proof.

We also investigated response-preserving orthogonal reconfiguration using a
matrix-exponential parameterization and compared it over $200$ matched starts
with the published implementation of the same problem. The published method
performs better on its reference benchmarks, with the difference increasing
with the number of internal resonators, and neither method reaches the
sixteenth-order Example-B topology in any start. The rotation route remains
useful as an initializer: followed by a hard-masked polish, it produces a
finished design from $62$ of $80$ starts at that order. Across the tested
benchmarks, the direct and rotation routes do not show a statistically
resolved difference in completed-design success rates. We therefore recommend
the direct route primarily because it avoids the unconstrained fit, rotation
parameterization, and separate rotation stage.

The end-to-end front end further shows that a conventional multiband
specification can be converted directly into a coupling matrix. For the
fabricated dual-band example, the specified band edges, transmission zeros,
and return-loss requirement admit a folded realization with one fewer
cross-coupling. By contrast, no such realization was found in $40$ constrained
starts for the response produced by the published rounded coefficients; the
best constrained result had $\max|\Delta S|=0.416$. This distinction is
important because the published rounded matrix does not reproduce the
idealized equiripple specification exactly.

Finally, the differentiable forward model allows the synthesized design to be
re-centered for tolerance robustness without deriving new sensitivities. For
the nominal design studied here, differentiable Monte Carlo centering increases
the model-predicted yield from $50.9\%$ to $67.0\pm0.5\%$ under the additive
coupling-error model. The centered design does so while reducing nominal
worst-case return-loss margin and reactivating the cross-coupling eliminated by
the nominal synthesis. The preferred realization therefore depends on the
fabrication-error model and objective used; this conclusion is limited to the
two realizations and tolerance models examined here.

Section~\ref{sec:discussion} discusses several extensions, including implicit differentiation through the LM stationarity conditions, lossy and frequency-dependent forward models, and a differentiable electromagnetic surrogate. The three-parameter redundancy in Example~B also provides additional freedom to choose among response-equivalent coupling matrices, which could be useful for improving manufacturability.


\balance

\end{document}